\documentclass[12pt,english,american]{article}
\usepackage[T1]{fontenc}
\usepackage[latin9]{inputenc}
\usepackage{geometry}
\usepackage{color}
\usepackage{babel}
\usepackage{amsthm}
\usepackage{amstext}
\usepackage{amssymb}
\usepackage{mathrsfs}
\usepackage{amsmath}
\usepackage{graphicx}
\usepackage{cancel}   
\usepackage[unicode=true,pdfusetitle,
 bookmarks=true,bookmarksnumbered=false,bookmarksopen=false,
 breaklinks=true,pdfborder={0 0 0},backref=false,colorlinks=true]
 {hyperref}
 \usepackage{setspace}
\hypersetup{
 linkcolor=blue,citecolor=blue, urlcolor=blue}
\makeatletter

\def\qed{$\Box$\medskip}

\newcommand{\beq}{\begin{equation}}
\newcommand{\eeq}{\end{equation}}
\newcommand{\beqa}{\begin{eqnarray}}
\newcommand{\eeqa}{\end{eqnarray}}
\newcommand{\ben}{\begin{arabicenumerate}}
\newcommand{\een}{\end{arabicenumerate}}

\def\bel{\begin{lem} } 
\def\eel{\end{lem} }
\def\bet{\begin{thm}}
\def\eet{\end{thm}}
\def\bed{\begin{defn}}
\def\eed{\end{defn} }

\def\bec{\begin{cor}}
\def\eec{\end{cor}}
\def\ber{\begin{rem}}
\def\eer{\end{rem}}

\usepackage{enumitem}		
\theoremstyle{plain}
\newtheorem{thm}{\protect\theoremname}[section]
\theoremstyle{definition}
\newtheorem{defn}[thm]{\protect\definitionname}
\theoremstyle{plain}

\theoremstyle{plain}
\theoremstyle{remark}
\newtheorem{rem}[thm]{\protect\remarkname}
\theoremstyle{plain}
\newtheorem{lem}[thm]{\protect\lemmaname}
\theoremstyle{plain}
\newtheorem{cor}[thm]{\protect\corollaryname}

\usepackage{txfonts,refstyle,xcolor}

\newref{con}{name = Conjecture\ }
\newref{prop}{name = Proposition\ }
\newref{def}{name = Definition\ }
\newref{sec}{name = Section\ }
\newref{sub}{name = Section\ }
\newref{thm}{name = Theorem\ }
\newref{lem}{name = Lemma\ }
\newref{cor}{name = Corollary\ }
\newref{fig}{name = Figure\ }
\newref{rem}{name =  Remark\ }

\usepackage{bbm}
\newcommand{\charf}{\mathbbm{1}}

\usepackage[all]{xy}

\newcommand{\xyR}[1]{%
     \makeatletter
     \xydef@\xymatrixrowsep@{#1}
     \makeatother
}

\newcommand{\xyC}[1]{%
     \makeatletter
     \xydef@\xymatrixcolsep@{#1}
     \makeatother
}

\newcommand{\ncol}[1]{\color{normalcolor}}

\makeatother

\addto\captionsamerican{\renewcommand{\corollaryname}{Corollary}}
\addto\captionsamerican{\renewcommand{\definitionname}{Definition}}
\addto\captionsamerican{\renewcommand{\lemmaname}{Lemma}}
\addto\captionsamerican{\renewcommand{\propositionname}{Proposition}}
\addto\captionsamerican{\renewcommand{\remarkname}{Remark}}
\addto\captionsamerican{\renewcommand{\theoremname}{Theorem}}
\addto\captionsenglish{\renewcommand{\corollaryname}{Corollary}}
\addto\captionsenglish{\renewcommand{\definitionname}{Definition}}
\addto\captionsenglish{\renewcommand{\lemmaname}{Lemma}}
\addto\captionsenglish{\renewcommand{\propositionname}{Proposition}}
\addto\captionsenglish{\renewcommand{\remarkname}{Remark}}
\addto\captionsenglish{\renewcommand{\theoremname}{Theorem}}
\providecommand{\corollaryname}{Corollary}
\providecommand{\definitionname}{Definition}
\providecommand{\lemmaname}{Lemma}
\providecommand{\propositionname}{Proposition}
\providecommand{\remarkname}{Remark}
\providecommand{\theoremname}{Theorem}

\addto\captionsamerican{\renewcommand{\corollaryname}{Corollary}}
\addto\captionsamerican{\renewcommand{\definitionname}{Definition}}
\addto\captionsamerican{\renewcommand{\lemmaname}{Lemma}}
\addto\captionsamerican{\renewcommand{\propositionname}{Proposition}}
\addto\captionsamerican{\renewcommand{\remarkname}{Remark}}
\addto\captionsamerican{\renewcommand{\theoremname}{Theorem}}
\addto\captionsenglish{\renewcommand{\corollaryname}{Corollary}}
\addto\captionsenglish{\renewcommand{\definitionname}{Definition}}
\addto\captionsenglish{\renewcommand{\lemmaname}{Lemma}}
\addto\captionsenglish{\renewcommand{\propositionname}{Proposition}}
\addto\captionsenglish{\renewcommand{\remarkname}{Remark}}
\addto\captionsenglish{\renewcommand{\theoremname}{Theorem}}
\providecommand{\corollaryname}{Corollary}
\providecommand{\definitionname}{Definition}
\providecommand{\lemmaname}{Lemma}
\providecommand{\propositionname}{Proposition}
\providecommand{\remarkname}{Remark}
\providecommand{\theoremname}{Theorem}

\begin{document}
\title{Lie-Schwinger block-diagonalization and gapped quantum chains} 
  \author{J. Fr\"ohlich \footnote{email: juerg@phys.ethz.ch}\\
 Institut f\"ur Theoretiche Physik, ETH-Z\"urich\\
 Z\"urich, Switzerland\\
A. Pizzo \footnote{email: pizzo@mat.uniroma2.it}\\
Dipartimento di Matematica, Universit\`a di Roma ``Tor Vergata",\\
 Roma, Italy}

\date{24/02/2019}

\maketitle

\abstract{We study quantum chains whose Hamiltonians are perturbations by bounded interactions of short range of a Hamiltonian that does not couple the degrees of freedom located at different sites of the chain and has a strictly positive energy gap above its ground-state energy. We prove that, for small values of a coupling constant, the spectral gap of the perturbed Hamiltonian above its ground-state energy  is bounded from below by a positive constant \textit{uniformly} in the length of the chain. In our proof we use a novel method based on \emph{local} Lie-Schwinger conjugations of the Hamiltonians associated with connected subsets of the chain.  }
\\
\section{Introduction: Models and Results}
In this paper, we study spectral properties of Hamiltonians of some family of quantum chains with bounded interactions of short range, including the \textit{Kitaev chain}, \cite{GST}, \cite{KST}. We are primarily interested in determining the multiplicity of the ground-state energy and in estimating the size of the spectral gap above the ground-state energy of Hamiltonians of such chains, as the length of the chains tend to infinity. We will consider a family of Hamiltonians for which we will prove that their ground-state energy is finitely degenerate and the spectral gap above the ground-state energy  is bounded from below by a positive constant, \textit{uniformly} in the length of the chain.
Connected sets of Hamiltonians with these properties represent what people tend to call (somewhat misleadingly) a ``topological phase''. Our analysis is motivated by recent wide-spread interest in characterising topological phases of matter;
see, e.g., \cite{MN}, \cite{NSY}, \cite{BN}.

Results similar to the ones established in this paper have been proven before, often using so-called ``cluster expansions''; see \cite{DFF}, \cite{FFU}, \cite{KT}, \cite{Y}, \cite{KU}, \cite{DS} \cite{H} and refs. given there. The purpose of this paper is to introduce a novel method to analyse spectral properties of Hamiltonians of quantum chains near their ground-state energies. This method is based on \textit{iterative unitary conjugations} of the Hamiltonians, which serve to block-diagonalise them with respect to a fixed orthogonal projection 
and its orthogonal complement; (see \cite{DFFR} for similar ideas in a simpler context). Ideas somewhat similar to those presented in this paper
have been used in work of J. Z. Imbrie, \cite{I1}, \cite{I2}.

\subsection{A concrete family of quantum chains}
The Hilbert space of pure state vectors of the quantum chains studied in this paper has the form
\begin{equation}\label{tensorprod}
\mathcal{H}^{(N)}:= \bigotimes_{j=1}^{N} \mathcal{H}_{j}\,,
\end{equation}
where $\mathcal{H}_{j}\simeq \mathbb{C}^{M}, \, \forall j=1,2,\dots,$ and where $M$ is an arbitrary, but $N$-independent finite integer. Let $H$ be a positive $M\times M$ matrix with the properties that $0$ is an eigenvalue of $H$ corresponding to an eigenvector $\Omega \in \mathbb{C}^{M}$, and 
$$H\vert_{\lbrace \mathbb{C} \Omega \rbrace^{\perp}} \geq \charf \,.$$
We define
\begin{equation}\label{H_i}
H_{i}:= \charf_{1}\otimes \dots \otimes \underset{\underset{i^{th} \text{slot}}{\uparrow}}{H} \otimes \dots \charf_{N}\,.
\end{equation}
By $P_{\Omega_i}$ we denote the orthogonal projection onto the subspace
\begin{equation}\label{vacuum_i}
\mathcal{H}_{1}\otimes \dots \otimes \underset{\underset{i^{th} \text{slot}}{\uparrow}}{\lbrace \mathbb{C} \Omega \rbrace}\otimes \dots \otimes \mathcal{H}_{N} \subset \mathcal{H}^{(N)}\,, \quad \text{  and}\quad   P_{\Omega_i}^{\perp} := \charf - P_{\Omega_i}\,.
\end{equation}
Then 
\begin{equation*} 
H_i = P_{\Omega_i} H_i P_{\Omega_i} + P_{\Omega_i}^{\perp} H_i P_{\Omega_i}^{\perp} \,,
\end{equation*}
with
\begin{equation}\label{gaps}
P_{\Omega_i} H_i P_{\Omega_i}=0\,,\quad P_{\Omega_i}^{\perp} H_i P_{\Omega_i}^{\perp} \geq P_{\Omega_i}^{\perp}\,.
\end{equation}
We study quantum chains on the graph $I_{N-1;1}:= \lbrace 1, \dots, N\rbrace, \,N< \infty$ arbitrary, with a Hamiltonian of the form
\begin{equation}\label{Hamiltonian}
K_{N}\equiv K_{N}(t):= \sum_{i=1}^{N} H_i + t \sum_{\underset{k \leq \bar{k}}{I_{k;i}\subset I_{N-1;1}}} V_{I_{k;i}}\,, 
\end{equation}
where $\bar{k} < \infty$ is an arbitrary, but fixed integer,
$I_{k;i}$ is the ``interval'' given by $\lbrace i, \dots, i+k\rbrace, \, i=1, \dots, N-k$, \, and $V_{I_{k;i}}$ is a symmetric matrix acting on $\mathcal{H}^{(N)}$ with the property that 
\begin{equation}\label{potential}
V_{I_{k;i}} \,\, \text{  acts as the identity on  }\,\, \bigotimes_{j\in I_{N-1;1}\,,\, j\notin I_{k;i}} \mathcal{H}_{j}\,,
\end{equation}
and $t\in \mathbb{R}$ is a coupling constant. (We call $I_{k;i}$ the ``support'' of $V_{I_{k;i}}$.) Without loss of generality, we may assume that
\begin{equation}\label{norms}
\Vert V_{I_{k;i}} \Vert \leq 1\,.
\end{equation}
A concrete example of a quantum chain we are able to analyse is the (generalised) ``Kitaev chain'', which has a Hamiltonian that is a small perturbation of the following quadratic Hamiltonian:
\begin{equation}\label{Kitaev}
H_N := -\mu \sum_{J=1}^{N} c_{j}^{\dagger} c_{j}\, -\, \sum_{j=1}^{N-1} \big(\tau c_{j}^{\dagger} c_{j+1} + h.c. + \Delta c_{j}c_{j+1} + h.c. \big)\,,
\end{equation}
where $c_{j}^{\dagger}, c_{j}, \, j=1,\dots, N,$ are fermi creation- and annihilation operators satisfying canonical anti-commutation relations, $\mu$ is a chemical potential, $\tau$ is a hopping amplitude, and $\Delta$ is a pairing amplitude. Using appropriate linear combinations of the operators $c_{j}^{\dagger}, c_j$, the Hamiltonian $H_N$, as well as certain small perturbations thereof, can be cast in the form given in Eq. (\ref{Hamiltonian}). See Sect. 4 for details.

Another example of a quantum chain that can be treated with the methods of this paper is an anisotropic Heisenberg chain corresponding to a small quantum perturbation of the ferromagnetic Ising chain, with domain walls interpreted as the elementary finite-energy excitations of the ferromagnetically ordered ground-state of the chain. A detailed analysis of such examples, as well as examples where the dimension, $M$, of the Hilbert spaces $\mathcal{H}_j$ is infinite is deferred to another paper.

\subsection{Main result}
The main result in this paper is the following theorem proven in Section \ref{proofs}, (see Theorem \ref{main-res}).

{\bf{Theorem.}}
\textit{Under the assumption that (\ref{gaps}), (\ref{potential}) and (\ref{norms}) hold, the Hamiltonian $K_{N}$ defined in (\ref{Hamiltonian}) has the following properties: There exists some $t_0 > 0$ such that, for any $t\in \mathbb{R}$ with 
$\vert t \vert < t_0$, and for all $N < \infty$,
\begin{enumerate}
\item[(i)]{ $K_{N}$ has a unique ground-state; and}
\item[(ii)]{ the energy spectrum of $K_N$ has a strictly positive gap, $\Delta_{N}(t) \geq \frac{1}{2}$, above the ground-state energy.}
\end{enumerate}
}

\begin{rem}
The ground-state of $K_{N}$ may depend on ``boundary conditions'' at the two ends of the chain, in which case several different ground-states may exist. A simple example of this phenomenon is furnished by the anisotropic Heisenberg chain described above, with $+$ or $-$ boundary conditions imposed at the ends of the chain.
\end{rem}

Results similar to the theorem stated above have appeared in the literature; see, e.g., \cite{DS}. The main novelty introduced in this paper is our method of proof.

We define
\begin{equation}\label{vacuum-proj}
P_{vac}:=\bigotimes_{i=1}^{N} P_{\Omega_i}\,.
\end{equation}
Note that $P_{vac}$ is the orthogonal projection onto the ground-state of the operator $K_{N}(t=0)=\sum_{i=1}^{N} H_{i}$. Our aim is to find an anti-symmetric matrix $S_{N}(t)=-S_{N}(t)^{\dagger}$ acting on $\mathcal{H}^{(N)}$ (so that 
exp$\big(\pm S_{N}(t)\big)$ is unitary) with the property that, after conjugation, the operator
\begin{equation}\label{conjug}
e^{S_{N}(t)}K_{N}(t)e^{-S_{N}(t)}=: \widetilde{K}_{N}(t)
\end{equation}
is \textit{``block-diagonal''} with respect to $P_{vac}$, $P_{vac}^{\perp} (:= \charf - P_{vac})$, in the sense that $P_{vac}$ projects onto the ground-state of $\widetilde{K}_{N}(t)$,
\begin{equation}\label{block-diag}
\widetilde{K}_{N}(t)= P_{vac} \widetilde{K}_{N}(t) P_{vac} + P_{vac}^{\perp} \widetilde{K}_{N}(t) P_{vac}^{\perp}\,,
\end{equation}
and 
\begin{equation}\label{gapss}
\text{infspec}\left(P_{vac}^{\perp}\widetilde{K}_{N}(t) P_{vac}^{\perp} \vert_{P_{vac}^{\perp} \mathcal{H}^{(N)}}\right)
\geq \text{infspec} \left(P_{vac} \widetilde{K}_{N}(t) P_{vac} \vert_{P_{vac}\mathcal{H}^{(N)}}\right) + \Delta_{N}(t)\,,
\end{equation}
with $\Delta_{N}(t) \geq \frac{1}{2}$, for $\vert t \vert < t_0$, \textit{uniformly} in $N$.
The iterative construction of the operator $S_{N}(t)$, yielding (\ref{block-diag}), and the proof of (\ref{gapss}) are the main tasks to be carried out. Formal aspects of our construction are described in Sect. 2. In Sect. 3, the proof of convergence of our construction of the operator $S_{N}(t)$ and the proof of a lower bound on the spectral gap $\Delta_{N}(t)$, for sufficiently small values of $\vert t \vert$, are presented, with a few technicalities deferred to Appendix A. In Sect. 4, the example of the (generalized) Kitaev chain is studied.
\\


{\bf{Notation}}
\\

\noindent
1) Notice that $I_{k;q}$  can also be seen as a connected one-dimensional graph with $k$ edges connecting  the $k+1$ vertices $q,1+q,\dots, k+q$, or as an ``interval'' of length $k$ whose left end-point coincides with $q$. 
\\

\noindent
2) We use the same symbol for the operator $O_j$ acting on $\mathcal{H}_j$ and the corresponding operator $$\charf_{i}\otimes\dots \otimes  \charf_{j-1}\otimes O_j \otimes \charf_{j+1}\dots \otimes \charf_l$$ acting on $ \bigotimes_{k=i}^{l} \mathcal{H}_k$, for any $i\leq j\leq l$.
\\

{\bf{Acknowledgements.}}
A.P.  thanks  the Pauli Center, Z\"urich, for hospitality in Spring 2017 when this project got started. A.P. also acknowledges the MIUR Excellence Department Project awarded to the Department of Mathematics, University of Rome Tor Vergata, CUP E83C18000100006.

\setcounter{equation}{0}
\section{\emph{Local} conjugations based on Lie-Schwinger series}

In this section we describe some of the key ideas underlying our proof of the theorem announced in the previous section.
We study quantum chains with Hamiltonians $K_{N}(t)$ of the form described in (\ref{Hamiltonian}) acting on the Hilbert space $\mathcal{H}^{(N)}$ defined in (\ref{tensorprod}).
As announced in Sect. 1, our aim is to block-diagonalize $K_{N}(t)$, for $\vert t \vert$ small enough, by conjugating it by a sequence of unitary operators chosen according to the ``Lie-Schwinger procedure'' (supported on subsets of $\lbrace 1, \dots, N \rbrace$ of successive sites). The block-diagonalization will concern operators acting on tensor-product  spaces of the sort 
$\mathcal{H}_{q} \otimes \dots \otimes \mathcal{H}_{k+q}$ (and acting trivially on the remaining tensor factors), and it will be with respect to the projection onto the ground-state (``vacuum'') subspace, $\lbrace \mathbb{C}(\Omega_{q}\otimes \dots \otimes \Omega_{k+q})\rbrace$, contained in $\mathcal{H}_{q} \otimes \dots \otimes \mathcal{H}_{k+q}$ and its orthogonal complement. Along the way, new interaction terms are being created  whose support corresponds to ever longer intervals (connected subsets) of the chain.

\subsection{Block-diagonalization: Definitions and formal aspects}
For each $k$, we consider $(N-k)$ block-diagonalization steps, each of them associated with  a subset $I_{k;q},\, q=1, \dots, N-k$. 
The block-diagonalization of the Hamiltonian will be  with respect to the subspaces associated with the projectors in (\ref{pro-minus})-(\ref{pro-plus}), introduced below.
By $(k,q)$ we label the block-diagonalization step associated with $I_{k;q}$. We introduce an ordering amongst these steps:
\begin{equation}
(k',q') \succ (k,q)
\end{equation} 
if $k'> k$ or if $k'=k$ and $q'>q$. 

\noindent
Our original Hamiltonian is denoted by $K^{(0,N)}_N :=K_{N}(t)$. We proceed to the first block-diagonalisation step yielding $K_{N}^{(1,1)}$. The index $(0,N)$  is our initial choice of the index $(k,q)$: all the on-site terms in the Hamiltonian, i.e, the terms $H_i$, are block-diagonal with respect to the subspaces associated with the projectors in (\ref{pro-minus})-(\ref{pro-plus}), for $l=0$.
Our goal is to arrive at a Hamiltonian of the form
\begin{eqnarray}\label{kappa-k-q}
K_N^{(k,q)}
& :=&\sum_{i=1}^{N}H_{i}+t\sum_{i=1}^{N-1}V^{(k,q)}_{I_{1;i}}+t\sum_{i=1}^{N-2}V^{(k,q)}_{I_{2;i}}+\dots+t\sum_{i=1}^{N-k}V^{(k,q)}_{I_{k;i}} \label{def-transf-ham}\\
& &+t\sum_{i=1}^{N-k-1}V^{(k,q)}_{I_{k+1;i}}+\dots+t\sum_{i=1}^{2}V^{(k,q)}_{I_{N-2;i}}+tV^{(k,q)}_{I_{N-1;1}}\label{def-transf-ham-bis}
\end{eqnarray}
after the block-diagonalization step $(k,q)$,
with the following properties:
\begin{enumerate}
\item
For a fixed $I_{l;i}$,  the corresponding potential term changes, at each step of the block-diagonalization procedure, up to the step $(k,q)\equiv (l,i)$; hence $V^{(k,q)}_{I_{l;i}}$ is the potential term associated with the interval $I_{l;i}$ at step $(k,q)$ of the block-diagonalization, and the superscript $(k,q)$ keeps track of the changes in the potential term in step $(k,q)$. The operator $V^{(k,q)}_{I_{l;i}}$ acts as the identity on the spaces $\mathcal{H}_j$ for $j\neq i,i+1,\dots,i+l$; the description of how these terms are created and estimates on their norms are deferred to Section \ref{proofs};
\item
for all sets $I_{l;i}$ with $ (l,i)\prec (k,q)$ and for the set $I_{l;i} \equiv I_{k;q} $, the associated potential $V^{(k,q)}_{I_{l; i}}$ is block-diagonal w.r.t. the decomposition of the identity into the sum of projectors
\begin{equation}\label{pro-minus}
P^{(-)}_{I_{l;i}}:= P_{\Omega_{i}}\otimes P_{\Omega_{i+1}}\otimes \dots \otimes P_{\Omega_{i+l}}\,,
\end{equation}
\begin{equation}\label{pro-plus}
P^{(+)}_{I_{l;i}}:= (P_{\Omega_{i}}\otimes P_{\Omega_{i+1}}\otimes \dots \otimes P_{\Omega_{i+l}})^{\perp}\,.
\end{equation}
\end{enumerate}

\begin{rem}\label{remark-decomp}
It is important to notice that if $V^{(k,q)}_{I_{l;i}}$ is block-diagonal w.r.t. the decomposition of the identity into $$P^{(+)}_{I_{l;i}}+P^{(-)}_{I_{l;i}}\,,$$
i.e., $$V^{(k,q)}_{I_{l;i}}=P^{(+)}_{I_{l;i}}V^{(k,q)}_{I_{l;i}}P^{(+)}_{I_{l;i}}+P^{(-)}_{I_{l;i}}V^{(k,q)}_{I_{l;i}}P^{(-)}_{I_{l;i}}\,\,,$$  then, for $ I_{l;i} \subset I_{r;j}$,
we have that
$$P^{(+)}_{I_{r;j}}\Big[P^{(+)}_{I_{l;i}}V^{(k,q)}_{I_{l;i}}P^{(+)}_{I_{l;i}}+P^{(-)}_{I_{l;i}}V^{(k,q)}_{I_{l;i}}P^{(-)}_{I_{l;i}}\Big]P^{(-)}_{I_{r;j}}=0\,.$$
To see that the first term vanishes, we use that
\begin{equation}
P^{(+)}_{I_{l;i}}\,P^{(-)}_{I_{r;j}}=0\,,
\end{equation}
while, in the second term, we use that
\begin{equation}
P^{(-)}_{I_{l;i}}V^{(k,q)}_{I_{l;i}}P^{(-)}_{I_{l;i}}\,P^{(-)}_{I_{r;j}}=P^{(-)}_{I_{r;j}}P^{(-)}_{I_{l;i}}V^{(k,q)}_{I_{l;i}}P^{(-)}_{I_{l;i}}P^{(-)}_{I_{r;j}}
\end{equation}
and
\begin{equation}
P^{(+)}_{I_{r;j}}P^{(-)}_{I_{r;j}}=0\,.
\end{equation}

\noindent
Hence $V^{(k,q)}_{I_{l;i}}$ is also block-diagonal with respect to the decomposition of the identity into  $$P^{(+)}_{I_{r;j}}+P^{(-)}_{I_{r;j}}\,.$$
However, notice that
\begin{equation}
P^{(-)}_{I_{r;j}}\Big[P^{(+)}_{I_{l;i}}V^{(k,q)}_{I_{l;i}}P^{(+)}_{I_{l;i}}+P^{(-)}_{I_{l;i}}V^{(k,q)}_{I_{l;i}}P^{(-)}_{I_{l;i}}\Big]P^{(-)}_{I_{r;j}}
=P^{(-)}_{I_{r;j}}\,V^{(k,q)}_{I_{l;i}}\,P^{(-)}_{I_{r;j}}
\end{equation}
but 
$$P^{(+)}_{I_{r;j}}\Big[P^{(+)}_{I_{l;i}}V^{(k,q)}_{I_{l;i}}P^{(+)}_{I_{l;i}}+P^{(-)}_{I_{l;i}}V^{(k,q)}_{I_{l;i}}P^{(-)}_{I_{l;i}}\Big]P^{(+)}_{I_{r;j}}$$
remains as it is.
\end{rem}
\begin{rem}\label{rem-block}
The block-diagonalization procedure that we will implement enjoys the property that the terms  block-diagonalized along the process do not change, anymore, in subsequent steps.
\\

\end{rem}
\subsection{Lie-Schwinger conjugation associated with $I_{k;q}$}
Here we explain the block-diagonalization procedure from $(k,q-1)$ to $(k,q)$ by which the term $V^{(k,q-1)}_{I_{k;q}}$ is transformed  to a new operator, $V^{(k,q)}_{I_{k;q}}$, which is block-diagonal  w.r.t.  the decomposition of the identity into  $$P^{(+)}_{I_{k;q}}+P^{(-)}_{I_{k;q}}\,.$$
We note that the steps of the type\footnote{The initial step, $(0,N)\rightarrow (1,1)$,  is of this type; see the definitions in (\ref{initial-V}) corresponding to a Hamiltonian $K_N$ with \emph{nearest-neighbor interactions}.} $(k, N-k)\, \rightarrow \,(k+1, 1)$ are somewhat different,  because the first index (i.e., the number of edges of the interval) is changing from $k$ to $k+1$. Hence  we start by showing how our procedure works for them. Later we deal with general steps $(k, q-1)\, \rightarrow \,(k, q)$, with $N-k\geq q\geq 2$. 

\noindent
We recall that the Hamiltonian $K_N^{(k,N-k)}$ is given by
\begin{eqnarray}
K_N^{(k,N-k)}
& :=&\sum_{i=1}^{N}H_{i}+t\sum_{i=1}^{N-1}V^{(k,N-k)}_{I_{1;i}}+t\sum_{i=1}^{N-2}V^{(k,N-k)}_{I_{2;i}}+\dots+t\sum_{i=1}^{N-k}V^{(k,N-k)}_{I_{k;i}}\\
& &+t\sum_{i=1}^{N-k-1}V^{(k,N-k)}_{I_{k+1;i}}+\dots+t\sum_{i=1}^{2}V^{(k,N-k)}_{I_{N-2;i}}+tV^{(k,N-k)}_{I_{N-1;1}}
\end{eqnarray}
and has the following properties
\begin{enumerate}
\item
each operator $V^{(k,N-k)}_{I_{l;i}}$ acts as the identity on the spaces $\mathcal{H}_j$ for $j\neq i,i+1,\dots,i+l$. In Section \ref{proofs} we explain how these terms are created and their norms estimated;
\item
each operator $V^{(k,N-k)}_{I_{l;i}}$, with $l\leq k$,  is block-diagonal w.r.t. the decomposition of the identity into the sum of projectors in (\ref{pro-minus})-(\ref{pro-plus}).
\end{enumerate}

With the next block-diagonalization step, labeled by $(k+1,1)$, we want to block-diagonalize the interaction term $V^{(k,N-k)}_{I_{k+1;1}}$, considering the operator
\begin{equation}
G_{I_{k+1;1}}:=\sum_{i\subset I_{k+1;1} }H_i+t\sum_{I_{1;i} \subset I_{k+1;1}} V^{(k, N-k)}_{I_{1;i}}+\dots+t\sum_{I_{k;i}\subset I_{k+1;1}}V^{(k, N-k)}_{I_{k;i}}
\end{equation}
as the ``unperturbed" Hamiltonian. This operator is block-diagonal w.r.t. the decomposition of  the identity in  (\ref{decomp-id}), i.e., 
\begin{equation}
G_{I_{k+1;1}}=P^{(+)}_{I_{k+1;1}}G_{I_{k+1;1}}P^{(+)}_{I_{k+1;1}}+P^{(-)}_{I_{k+1;1}}G_{I_{k+1;1}}P^{(-)}_{I_{k+1;1}}\,;
\end{equation}
see Remarks \ref{remark-decomp} and \ref{rem-block}.
We also define
\begin{equation}
E_{I_{k+1;1}}:=\inf\text{spec} \, G_{I_{k+1;1}}
\end{equation}
and we temporarily assume that $$G_{I_{k+1;1}}P^{(-)}_{I_{k+1;1}}=E_{I_{k+1;1}}P^{(-)}_{I_{k+1;1}}\,.$$
 Next, we sketch a convenient formalism used to construct our block-diagonalisation operations, below; (for further details the reader is referred to Sects. 2 and 3 of \cite{DFFR}). We define
\begin{equation}
ad\, A\,(B):=[A\,,\,B]\,,
\end{equation}
where $A$ and $B$ are bounded operators,
and, for $n\geq 2$,
\begin{equation}
ad^n A\,(B):=[A\,,\,ad^{n-1} A\,(B)]\,.
\end{equation}
In the block-diagonalization step $(k+1,1)$, we use the operator
\begin{equation}\label{unitary-k+1}
U_{I_{k+1,1}}:=\,e^{-S_{I_{k+1;1}}}\,,
\end{equation}
with
\begin{equation}\label{formula-S}
S_{I_{k+1;1}}:=\sum_{j=1}^{\infty}t^j(S_{I_{k+1;1}})_j\,,
\end{equation}
where
\begin{itemize}
\item
\begin{equation}\label{def-S}
(S_{I_{k+1;1}})_j:=ad^{-1}\,G_{I_{k+1;1}}\,((V^{(k, N-k)}_{I_{k+1;1}})^{od}_j):=\frac{1}{G_{I_{k+1;1}}-E_{I_{k+1;1}}}P^{(+)}_{I_{k+1;1}}\,(V^{(k, N-k)}_{I_{k+1;1}})_j\,P^{(-)}_{I_{k+1;1}}-h.c.\,,
\end{equation}
where \emph{od} means ``off-diagonal" w.r.t. the decomposition of the identity into 
\begin{equation}\label{decomp-id}
P^{(+)}_{I_{k+1;1}}+P^{(-)}_{I_{k+1;1}}\,
\end{equation}
\item
$(V^{(k,N-k)}_{I_{k+1;1}})_1:=V^{(k,N-k)}_{I_{k+1;1}}$, and, for $j\geq 2$,
\begin{eqnarray}\label{formula-v_j}
& &(V^{(k,N-k)}_{I_{k+1;1}})_j:=\nonumber\\
& &\sum_{p\geq 2, r_1\geq 1 \dots, r_p\geq 1\, ; \, r_1+\dots+r_p=j}\frac{1}{p!}\text{ad}\,(S_{I_{k+1;1}})_{r_1}\Big(\text{ad}\,(S_{I_{k+1;1}})_{r_2}\dots (\text{ad}\,(S_{I_{k+1;1}})_{r_p}(G_{I_{k+1;1}})\dots \Big)\nonumber\\
& &+\sum_{p\geq 1, r_1\geq 1 \dots, r_p\geq 1\, ; \, r_1+\dots+r_p=j-1}\frac{1}{p!}\text{ad}\,(S_{I_{k+1;1}})_{r_1}\Big(\text{ad}\,(S_{I_{k+1;1}})_{r_2}\dots (\text{ad}\,(S_{I_{k+1;1}})_{r_p}(V^{(k,N-k)}_{I_{k+1;1}})\dots \Big)\,.\nonumber\\
\end{eqnarray}
\end{itemize}
We define
\begin{equation}\label{ham-k+1}
K_N^{(k+1,1)}:=U^{\dagger}_{I_{k+1,1}}\,K_N^{(k, N-k)}\,U_{I_{k+1,1}}\,, \,\,\text{   with  }\, U_{I_{k+1,1}} \, \text{  as in   }(\ref{unitary-k+1})\,.
\end{equation}
\\

After the block-diagonalization step labeled by $(k,q-1)$, with $q\leq N-k$, we obtain
\begin{eqnarray}
K_N^{(k,q-1)}
& :=&\sum_{i=1}^{N}H_{i}+t\sum_{i=1}^{N-1}V^{(k,q-1)}_{I_{1;i}}+t\sum_{i=1}^{N-2}V^{(k,q-1)}_{I_{2;i}}+\dots+t\sum_{i=1}^{N-k}V^{(k,q-1)}_{I_{k;i}} \\
& &+t\sum_{i=1}^{N-k-1}V^{(k,q-1)}_{I_{k+1;i}}+\dots+t\sum_{i=1}^{2}V^{(k,q-1)}_{I_{N-2;i}}+tV^{(k,q-1)}_{I_{N-1;1}}
\end{eqnarray}
where, for all sets $I_{k';q'}$, with $ (k'q') \prec (k,q-1)$, and for the set $I_{k;q-1}$, the associated $V^{(k,q-1)}_{I_{k';q'}}$ is block-diagonal.

\noindent
Next, in order to block-diagonalize the interaction term $V^{(k,q-1)}_{I_{k;q}}$, we conjugate the Hamiltonian with the operator
\begin{equation}
U_{I_{k;q}}:=\,e^{-S_{I_{k;q}}}\,,
\end{equation}
where
\begin{equation}
S_{I_{k;q}}:=\sum_{j=1}^{\infty}t^j(S_{I_{k;q}})_j\,,
\end{equation}
with
\begin{equation}\label{def-S-bis}
(S_{I_{k;q}})_j:=ad^{-1}\,G_{I_{k;q}}\,((V^{(k,q-1)}_{I_{k;q}})^{od}_j)\,
\end{equation}
and

\begin{equation}\label{def-G}
G_{I_{k;q}}:=\sum_{i\subset I_{k;q} }H_i+t\sum_{I_{1;i} \subset I_{k;q}} V^{(k, q-1)}_{I_{1;i}}+\dots+t\sum_{I_{k-1;i}\subset I_{k;q}}V^{(k, q-1)}_{I_{k-1;i}}\,;
\end{equation}

$$(V^{(k,q-1)}_{I_{k;q}})_1=V^{(k,q-1)}_{I_{k;q}}$$ 
and, for $j\geq 2$,\\
\vspace{0.2cm}

$(V^{(k,q-1)}_{I_{k;q}})_j\,:=$
\begin{eqnarray}
\quad& =&\sum_{p\geq 2, r_1\geq 1 \dots, r_p\geq 1\, ; \, r_1+\dots+r_p=j}\frac{1}{p!}\text{ad}\,(S_{I_{k;q}})_{r_1}\Big(\text{ad}\,(S_{I_{k;q}})_{r_2}\dots (\text{ad}\,(S_{I_{k;q}})_{r_p}(G_{I_{k;q}})\dots \Big)\nonumber \\
&+&\sum_{p\geq 1, r_1\geq 1 \dots, r_p\geq 1\, ; \, r_1+\dots+r_p=j-1}\frac{1}{p!}\text{ad}\,(S_{I_{k;q}})_{r_1}\Big(\text{ad}\,(S_{I_{k;q}})_{r_2}\dots (\text{ad}\,(S_{I_{k;q}})_{r_p}(V^{(k,q-1)}_{I_{k;q}})\dots \Big)\, \nonumber\\
\end{eqnarray}

We define
\begin{equation}\label{def-K}
K_N^{(k,q)}:=e^{S_{I_{k;q}}}\,K_N^{(k,q-1)}\,e^{-S_{I_{k;q}}}\,.
\end{equation}

\subsection{Gap of the local Hamiltonians $G_{I_{k;q}}$: Main argument}
In the rest of this section we outline the main arguments and estimates underlying our strategy. To simplify our presentation, we consider a \emph{nearest-neighbor interaction} with 
$${\|V_{I_{1;i}}\|=1}\,$$ 
and $t>0$ small enough. However, with obvious modifications, our proof can be adapted to general Hamiltonians of the type in (\ref{Hamiltonian}).
\\

\noindent
We assume that 
\begin{equation}\label{ass-2}
\|V^{(k,q-1)}_{I_{l;i}}\| \leq \frac{8 \cdot t^{\frac{l-1}{3}}}{(l+1)^2}\,.
\end{equation} 
(The number ``$8$'' does not have particular significance, but comes up in the inductive part of the proof of Theorem \ref{th-norms}.)

\noindent
We exibit the key mechanism underlying our method,  starting from the potential terms $V^{(k,q-1)}_{I_{1;i}}$. We already know that,  for any  $k>1$,  the operator $V^{(k,q-1)}_{I_{1;i}}$ is block-diagonalized, i.e., 
\begin{equation}\label{informal-in}
V^{(k,q-1)}_{I_{1;i}}=P^{(+)}_{I_{1;i}}V^{(k,q-1)}_{I_{1;i}}P^{(+)}_{I_{1;i}}+P^{(-)}_{I_{1;i}}V^{(k,q-1)}_{I_{1;i}}P^{(-)}_{I_{1;i}}\,.
\end{equation}
Hence  we can  write
\begin{eqnarray}
& &P^{(+)}_{I_{k;q}}\,\Big[\sum_{i\subset I_{k;q;} }H_i+t\sum_{I_{1;i} \subset I_{k;q}} V^{(k,q-1)}_{I_{1;i}}\Big]P^{(+)}_{I_{k;q}}\\
&=&P^{(+)}_{I_{k;q}}\,\Big[\sum_{i\subset I_{k;q} }H_i+t\sum_{I_{1;i} \subset I_{k;q}} P^{(+)}_{I_{1;i}}V^{(k,q-1)}_{I_{1;i}}P^{(+)}_{I_{1;i}}+t\sum_{I_{1;i} \subset I_{k;q}} P^{(-)}_{I_{1;i}}V^{(k,q-1)}_{I_{1;i}}P^{(-)}_{I_{1;i}}\Big]P^{(+)}_{I_{k;q}}\quad\quad\quad \label{1.57}
\end{eqnarray}
and observe that, by assumption (\ref{gaps}),
\begin{equation}\label{ineq-free}
\sum_{i\subset I_{k;q} }H_i \geq \sum_{i=q}^{k+q} P^{\perp}_{\Omega_i} \,.
\end{equation}
We will make use of a simple, but crucial inequality proven  in Corollary \ref{op-ineq-2}: For $1\leq l \leq L \leq N-r$, 
\begin{equation}\label{ineq-inter-0}
\sum_{i=l}^{L}P^{(+)}_{I_{r;i}}\leq (r+1) \sum_{i=l}^{L+r} P^{\perp}_{\Omega_i} \,.
\end{equation}
Due to assumption (\ref{ass-2}) and inequality (\ref{ineq-inter-0}), with $r=1$, $l=q$, $L=k+q-r$,  we have that
\begin{equation}\label{ineq-cor}
\sum_{I_{1;i} \subset I_{k;q}} P^{(+)}_{I_{1;i}}V^{(k,q-1)}_{I_{1;i}}P^{(+)}_{I_{1;i}}\leq 4\cdot \,\sum_{i=q}^{k+q} P^{\perp}_{\Omega_i}\,.
\end{equation}
Hence, recalling that $t>0$ and combining (\ref{ineq-free}) with (\ref{ineq-cor}), we conclude that
\begin{eqnarray}
(\ref{1.57})
&\geq & P^{(+)}_{I_{k;q}}\,\Big[(1- 4t)\,\sum_{i=q}^{k+q} P^{\perp}_{\Omega_i}\Big]P^{(+)}_{I_{k;q}}+ P^{(+)}_{I_{k;q}}\,\Big[t\sum_{I_{1;i} \subset I_{k;q}}P^{(-)}_{I_{1;i}} V^{(k,q-1)}_{I_{1;i}}P^{(-)}_{I_{1;i}}\Big]P^{(+)}_{I_{k;q}}\label{second-line}\\
&=& P^{(+)}_{I_{k;q}}\,\Big[(1- 4t)\sum_{i=q}^{k+q} P^{\perp}_{\Omega_i}\Big]P^{(+)}_{I_{k;q}}+ P^{(+)}_{I_{k;q}}\,\Big[t\sum_{I_{1;i} \subset I_{k;q}}\langle  V^{(k,q-1)}_{I_{1;i}} \rangle P^{(-)}_{I_{1;i}}\Big]P^{(+)}_{I_{k;q}}\,,\label{third-line}
\end{eqnarray}
where $$\langle  V^{(k, q-1)}_{I_{1;i}} \rangle:=\langle \Omega_{i}\otimes \Omega_{i+1}\,,\,V^{(k,q-1)}_{I_{1;i}} \,\Omega_{i}\otimes \Omega_{i+1} \rangle\,.$$
Next, substituting $P^{(-)}_{I_{1;i}}=\charf -P^{(+)}_{I_{1;i}}$ into (\ref{third-line}), we find that
\begin{eqnarray}
(\ref{1.57})
&\geq & P^{(+)}_{I_{k;q}}\,\Big[(1-4t)\sum_{i=q}^{k+q}P^{\perp}_{\Omega_i}-t\sum_{I_{1;i} \subset I_{k;q}}\langle  V^{(k,q-1)}_{I_{1;i}} \rangle P^{(+)}_{I_{1;i}}\Big]P^{(+)}_{I_{k;q}} \quad\quad\quad\quad  \label{fin-1}\\
& &+ P^{(+)}_{I_{k;q}}\,\Big[t\sum_{I_{1;i} \subset I_{k;q}}\langle  V^{(k,q-1)}_{I_{1;i}} \rangle \Big]P^{(+)}_{I_{k;q}}\\
&\geq&P^{(+)}_{I_{k;q}}\,\Big[(1-8t)\sum_{i=q}^{k+q}P^{\perp}_{\Omega_i}\Big]P^{(+)}_{I_{k;q}}\label{fin-2}\\
& &+ P^{(+)}_{I_{k;q}}\,\Big[t\sum_{I_{1;i} \subset I_{k;q}}\langle  V^{(k,q-1)}_{I_{1;i}} \rangle \Big]P^{(+)}_{I_{k;q}}\,,\label{informal-fin}
\end{eqnarray}
where, in the step from (\ref{fin-1}) to (\ref{fin-2}), we have used (\ref{ineq-cor}).
Iterating this argument yields the following lemma.
\begin{lem}\label{gap}
Assuming the bound in (\ref{ass-2}) and choosing $t$ so small that
\begin{equation}
1-8t-16t \sum_{l=3}^{k}l\frac{ t^{\frac{l-2}{3}}}{l^2}>0\,,
\end{equation}
the following inequality holds:
\begin{eqnarray}
P^{(+)}_{I_{k;q}}G_{I_{k;q}}P^{(+)}_{I_{k;q}}
&\geq&\Big(1-8t-16t \sum_{l=3}^{k}l\frac{ t^{\frac{l-2}{3}}}{l^2} \Big)\,\,P^{(+)}_{I_{k;q}}\nonumber \\
& &+ P^{(+)}_{I_{k;q}}\,\Big[t\sum_{I_{1; i} \subset I_{k;q}}\langle  V^{(k,q-1)}_{I_{1;i}} \rangle +\dots+t\sum_{I_{k-1;i}\subset  I_{k;q}}\langle V^{(k,q-1)}_{I_{k-1;i}}\rangle\Big]P^{(+)}_{I_{k;q}}\,.\label{final-eq-1}
\end{eqnarray}
\end{lem}

\noindent
\emph{Proof.}
This lemma serves to establish a bound on the spectral gap above the ground-state energy of the operator $G_{I_{k;q}}$.
Proceeding as in (\ref{informal-in})-(\ref{informal-fin}) we get that

\begin{eqnarray}
P^{(+)}_{I_{k;q}}G_{I_{k;q}}P^{(+)}_{I_{k;q}} 
&\geq &P^{(+)}_{I_{k;q}}\,\Big[ \Big(1-4t-8t \sum_{l=3}^{k}l\frac{ t^{\frac{l-2}{3}}}{l^2} \Big)\,\sum_{i=q}^{k+q}P^{\perp}_{\Omega_i}\Big]P^{(+)}_{I_{k;q}} \quad\quad\quad\quad \nonumber \\
& &+ P^{(+)}_{I_{k;q}}\,\Big[t\sum_{I_{1;i} \subset I_{k;q}}\langle  V^{(k,q-1)}_{I_{1;i}} \rangle  P^{(-)}_{I_{1;i}}+\dots+t\sum_{I_{k-1;i}\subset  I_{k;q}}\langle V^{(k,q-1)}_{I_{k-1;i}}\rangle P^{(-)}_{I_{k-1;i}} \Big]P^{(+)}_{I_{k;q}} \quad\quad\quad \\
&\geq &P^{(+)}_{I_{k;q}}\,\Big[  \Big(1-4t-8t \sum_{l=3}^{k}l\frac{ t^{\frac{l-2}{3}}}{l^2} \Big)\,\,\sum_{i=q}^{k+q} P^{\perp}_{\Omega_i}\Big]P^{(+)}_{I_{k;q}} \nonumber \\
& &+ P^{(+)}_{I_{k;q}}\,\Big[-t\sum_{I_{1;i} \subset I_{k;q}}\langle  V^{(k,q-1)}_{I_{1;i}} \rangle  P^{(+)}_{I_{1;i}}+\dots-t\sum_{I_{k-1;i}\subset  I_{k;q}}\langle V^{(k,q-1)}_{I_{k-1;i}}\rangle P^{(+)}_{I_{k-1;i}} \Big]P^{(+)}_{I_{k;q}} \nonumber \\
& &+ P^{(+)}_{I_{k;q}}\,\Big[t\sum_{I_{1;i} \subset I_{k;q}}\langle  V^{(k,q-1)}_{I_{1;i}} \rangle +\dots+t\sum_{I_{k-1;i} \subset  I_{k;q}}\langle V^{(k,q-1)}_{I_{k-1;i}}\rangle \Big]P^{(+)}_{I_{k;q}} \quad\quad\quad 
\end{eqnarray}
\begin{eqnarray}
\quad\quad\quad\quad\quad\quad &\geq &P^{(+)}_{I_{k;q}}\,\Big[  \Big(1-8t-16t\sum_{l=3}^{k}l\frac{ t^{\frac{l-2}{3}}}{l^2} \Big)\,\sum_{i=q}^{k+q} P^{\perp}_{\Omega_i}\Big]P^{(+)}_{I_{k;q}} \nonumber \\
& &+ P^{(+)}_{I_{k;q}}\,\Big[t\sum_{I_{1;i} \subset I_{k;q}}\langle  V^{(k,q-1)}_{I_{1;i}} \rangle +\dots+t\sum_{I_{k-1;i} \subset  I_{k;q}}\langle V^{(k,q-1)}_{I_{k-1;i}}\rangle \Big]P^{(+)}_{I_{k;q}} \quad\quad\quad \\
&\geq & \Big(1-8t-16t\sum_{l=3}^{k}l\frac{ t^{\frac{l-2}{3}}}{l^2} \Big)\,\,P^{(+)}_{I_{k;q}}+ P^{(+)}_{I_{k;q}}\,\Big[t\sum_{I_{1;i} \subset I_{k;q}}\langle  V^{(k,q-1)}_{I_{1;i}} \rangle +\dots+t\sum_{I_{k-1;i} \subset  I_{k;q}}\langle V^{(k,q-1)}_{I_{k-1;i}}\rangle \Big]P^{(+)}_{I_{k;q}}\nonumber
\end{eqnarray}
where Lemma \ref{op-ineq-1} is used in the last inequality.
\qed

Lemma \ref{gap} implies that, under  assumption (\ref{ass-2}),  the Hamiltonian $G_{I_{k;q}}$ has a spectral gap above its groundstate energy that can be estimated from below by $\frac{1}{2}$, for $t$ sufficiently small but \textit{independent} of $N$,  $k$, and $q$, as  stated in the Corollary below.
\begin{cor}\label{cor-gap}
For $t$ sufficiently small, but independent of $N$, $k$,  and $q$,  the Hamiltonian $G_{I_{k;q}}$  has  a spectral gap $\Delta_{I_{k;q}}\geq \frac{1}{2}$ above the ground-state energy. The ground-state of $G_{I_{k;q}}$ coincides with the ``vacuum'', $\bigotimes_{j\in I_{k;q}}\Omega_{j}$\,\,, in $\mathcal{H}_{I_{k;q}}$. We have the identity
\begin{eqnarray}
P^{(-)}_{I_{k;q}}G_{I_{k;q}}P^{(-)}_{I_{k;q}}
&= &P^{(-)}_{I_{k;q}}\, \Big[t\sum_{I_{1;i} \subset I_{k;q}}\langle  V^{(k,q-1)}_{I_{1;i}} \rangle P^{(-)}_{I_{1;i}} +\dots+t\sum_{I_{k-1;i}\subset  I_{k;q}}\langle V^{(k,q-1)}_{I_{k-1;i}}\rangle P^{(-)}_{I_{k-1;i}}  \Big]P^{(-)}_{I_{k;q}}\nonumber\\
&= &P^{(-)}_{I_{k;q}}\, \Big[t\sum_{ I_{1;i}\subset I_{k;q}}\langle  V^{(k,q-1)}_{I_{1;i}} \rangle+\dots+t\sum_{I_{k-1;i}\subset  I_{k;q}}\langle V^{(k,q-1)}_{I_{k-1;i}}\rangle \Big]P^{(-)}_{I_{k;q}}\,.\label{final-eq-2}
\end{eqnarray}

\end{cor}

\setcounter{equation}{0}
\section{An algorithm defining the operators $V^{(k,q)}_{I_{l;i}}$, and inductive control of block-diagonalization}\label{proofs}

Here we address the question of how the interaction terms evolve under our block-diagonaliza-\\tion steps. We propose to define and control an algorithm, $\alpha_{I_{k;q}}$, determining a map that sends each operator $V^{(k,q-1)}_{I_{l;i}}$ to a corresponding potential term supported on the same interval, but at the next block-diagonalization step, i.e., 

\begin{equation}
\alpha_{I_{k;q}}(V^{(k,q-1)}_{I_{l;i}})=:V^{(k,q)}_{I_{l;i}}\,.
\end{equation}
For this purpose,  it is helpful to study what happens to the interaction term 
$V^{(k,q-1)}_{I_{l;i}}$ after conjugation with $\text{exp}(S_{I_{k;q}})$, i.e., to consider the operator
\begin{equation}
e^{S_{I_{k;q}}}\,V^{(k,q-1)}_{I_{l;i}}\,e^{-S_{I_{k;q}}}\,,
\end{equation}
assuming that $S_{I_{k;q}}$ is well defined.
We start from $V_{I_{0;i}}^{(0,N)}:=H_i$ and follow the fate of these operators and the one of the potential terms.  As will follow from definition (\ref{def-int}), $V_{I_{0;i}}^{(k,q)}$ coincides with $H_i$,  for all $k$ and $q$. 
\\

\noindent
We distinguish four cases; (see Fig.\ref{fig:cases} for a graphical representation of the different cases):
\begin{itemize}
\item[1)] If $I_{l;i} \cap I_{k;q}=\emptyset$
then \begin{equation}\label{op-1}
e^{S_{I_{k;q}}}\,V^{(k,q-1)}_{I_{l;i}}\,e^{-S_{I_{k;q}}}=V^{(k,q-1)}_{I_{l;i}}\,,
\end{equation}
since $S_{I_{k;q}}$ acts as the identity on $\mathcal{H}_{I_{l;i}}:=\mathcal{H}_{i} \otimes \dots \otimes  \mathcal{H}_{i+l}$.
\item[2)] If $I_{k;q}\subset I_{l;i}$ then 
\begin{equation}\label{op-2}
e^{S_{I_{k;q}}}\,V^{(k,q-1)}_{I_{l;i}}\,e^{-S_{I_{k;q}}}=(V^{(k,q-1)}_{I_{l;i}})'\,\,,
\end{equation} 
where the right side  is an operator acting as the identity outside $\mathcal{H}_{I_{l;i}}$.
\item[3)] If $I_{l;i}\subseteq I_{k;q}$ then 
\begin{equation}
e^{S_{I_{k;q}}}\,V^{(k,q-1)}_{I_{l;i}}\,e^{-S_{I_{k;q}}}=(V^{(k,q-1)}_{I_{l;i}})'' \,\,,
\end{equation}
where the right side  is an operator acting as the identity outside $\mathcal{H}_{I_{k;q}}$.
\item[4)] If $I_{l;i}\cap I_{k;q}\neq \emptyset$\,, with $I_{l;i}\nsubseteq I_{k;q}$ and $I_{k;q}\nsubseteq I_{l;i}$\,, then  we use that 
\begin{equation}
e^{S_{I_{k;q}}}\,V^{(k,q-1)}_{I_{l;i}}\,e^{-S_{I_{k;q}}}=V^{(k,q-1)}_{I_{l;i}}+\sum_{n=1}^{\infty}\frac{1}{n!}\,ad^{n}S_{I_{k;q}}(V^{(k,q-1)}_{I_{l;i}}) \,.\label{growth-0}
\end{equation}
\end{itemize}
We will provide a precise definition of $V^{(k,q)}_{I_{l;i}}$ below; see Definition \ref{def-interections}. To prepare the grounds, some heuristic explanations may be helpful: Each operator $V^{(k,q)}_{I_{l;i}}$ can be thought of as resulting from the following operations:
\begin{itemize}
\item[I)]
A ``growth process", involving operators corresponding to shorter intervals, as described in point 4), above, by the terms on the very right side of (\ref{growth-0}).
\item[II)] Operations as in points 1) and 2), or as given by the first term on the right side of (\ref{growth-0}), which do not change the support of the operator (i.e, they do not change the length of the interval) and leave the norm of the operator invariant.
\item[III)] Operations as described in case 3), above, and made more explicit in the following remarks: By including all potentials\footnote{Recall that $V_{I_{0;i}}^{(0,N)}:=H_i$ and $V_{I_{0;i}}^{(k,q)}$ will coincide with $V_{I_{0;i}}^{(0,N)}$ for all $(k,q)$.} $V^{(k,q-1)}_{I_{l;i}}$, with $I_{l;i}\subset I_{k;q}$,  we obtain the operator denoted by $G_{I_{k;q}}$. Moreover, by construction of $S_{I_{k;q}}$,
\begin{equation}\label{conjugation}
e^{S_{I_{k;q}}}\,(G_{I_{k;q}}+tV^{(k,q-1)}_{I_{k;q}})\,e^{-S_{I_{k;q}}}=G_{I_{k;q}}+t\sum_{j=1}^{\infty}t^{j-1}(V^{(k,q-1)}_{I_{k;q}})^{diag}_j
\end{equation}
where \emph{``diag''} indicates that the corresponding operator is block-diagonal w.r.t. to the decomposition of the identity into $P^{(-)}_{I_{k;q}}+P^{(+)}_{I_{k;q}}$.
Hence: 
\begin{enumerate}
 \item[i)]
If $I_{l;i}\equiv I_{k;q}$ we set
\begin{equation}\label{diag-term}
V^{(k,q)}_{I_{l;i}\equiv I_{k;q}}:=\sum_{j=1}^{\infty}t^{j-1}(V^{(k,q-1)}_{I_{k;q}})^{diag}_j=e^{S_{I_{k;q}}}\,(\frac{G_{I_{k;q}}}{t}+V^{(k,q-1)}_{I_{k;q}})\,e^{-S_{I_{k;q}}}-\frac{G_{I_{k;q}}}{t}\,.
\end{equation} 
Clearly the operator $V^{(k,q)}_{I_{k;q}}$ acts as the identity outside $\mathcal{H}_{I_{k;q}}$ but in general $\|V^{(k,q)}_{I_{k;q}}\|\neq \|V^{(k,q-1)}_{I_{k;q}}\|$.

\item[ii)]
If $I_{l;i}\subset I_{k;q}$  we set
\begin{equation}\label{def-int}
V^{(k,q)}_{I_{l;i}}:=V^{(k,q-1)}_{I_{l;i}}\,,
\end{equation}
which is block-diagonal w.r.t. the decomposition of the identity into $P^{(+)}_{I_{k;q}}+P^{(-)}_{I_{k;q}}$, too, as explained in Remark \ref{remark-decomp}. Clearly the operator $V^{(k,q)}_{I_{l;i}}$ acts as the identity outside $\mathcal{H}_{I_{l;i}}$ and $\|V^{(k,q)}_{I_{l;i}}\|=\|V^{(k,q-1)}_{I_{l;i}}\|$\,.
\end{enumerate}
\begin{figure}
 \includegraphics[width=\linewidth]{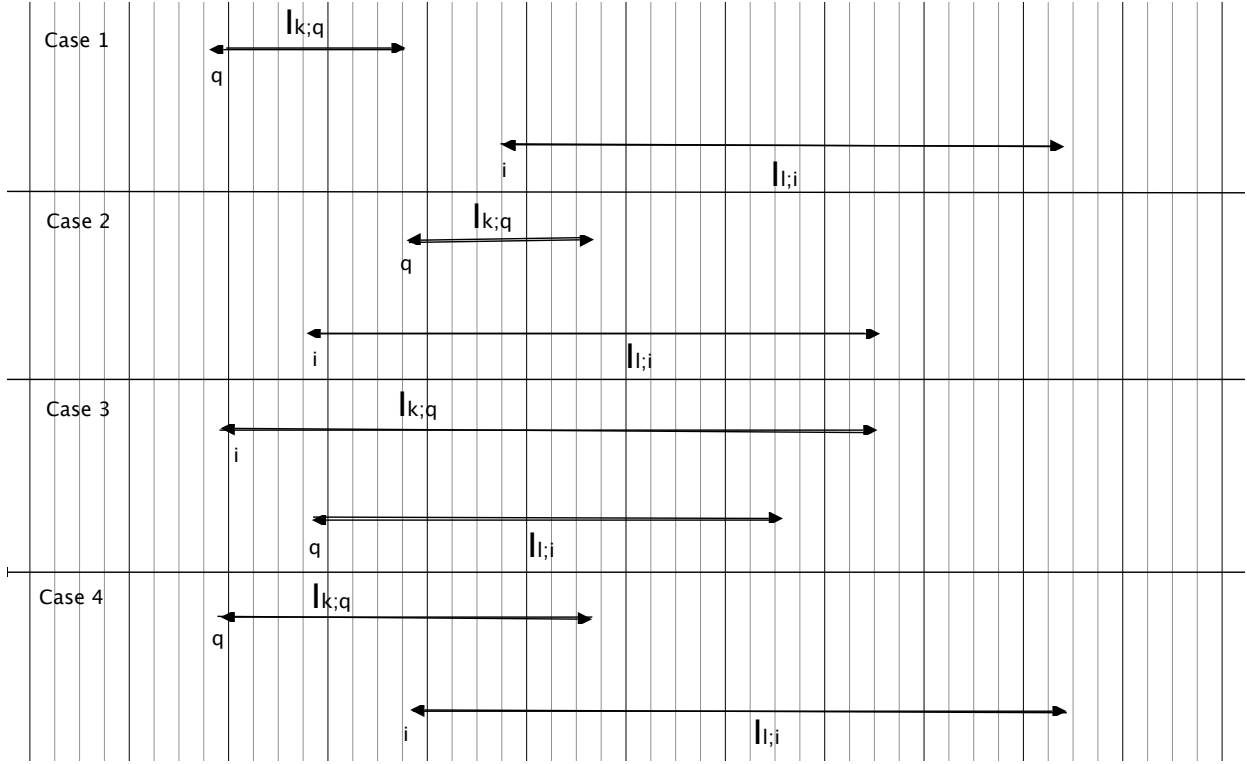}
 \caption{Relative positions of intervals $I_{k;q}$ and $I_{l;i}$}
 \label{fig:cases}
\end{figure}
Thus
the \underline{net result} of the conjugation  of the sum of the operators $V^{(k,q-1)}_{I_{l;i}}$  appearing on the left side of  eq. (\ref{conjugation})
can be re-interpreted as follows:

\noindent
a) The operators $V^{(k,q-1)}_{I_{l;i}}$, with $I_{l;i}\subset I_{k;q}$\,, are kept fixed in the step $(k,q-1)\rightarrow (k,q)$, i.e., we define $V^{(k,q)}_{I_{l;i}}:=V^{(k,q-1)}_{I_{l;i}}$,  hence
$$G_{I_{k;q}}=\sum_{i\subset I_{k;q} }H_i+t\sum_{I_{1;i} \subset I_{k;q}} V^{(k, q-1)}_{I_{1;i}}+\dots+t\sum_{I_{k-1;i}\subset I_{k;q}}V^{(k, q-1)}_{I_{k-1;i}}=\sum_{i\subset I_{k;q} }H_i+t\sum_{I_{1;i} \subset I_{k;q}} V^{(k, q)}_{I_{1;i}}+\dots+t\sum_{I_{k-1;i}\subset I_{k;q}}V^{(k, q)}_{I_{k-1;i}}$$

\noindent
b)  the operator $V^{(k,q-1)}_{I_{k;q}}$ is transformed to the operator $$V^{(k,q)}_{I_{k;q}}:=\sum_{j=1}^{\infty}t^{j-1}(V^{(k,q-1)}_{I_{k;q}})^{diag}_j$$  which is block-diagonal, and $$\|V^{(k,q)}_{I_{k;q}}\|\leq 2\|V^{(k,q-1)}_{I_{k;q}}\|\,,$$ as will be shown, assuming that $t>0$ is sufficiently small.
\end{itemize}

\subsection{The algorithm $\alpha_{I_{k;q}}$}
\begin{figure}
 \includegraphics[width=\linewidth]{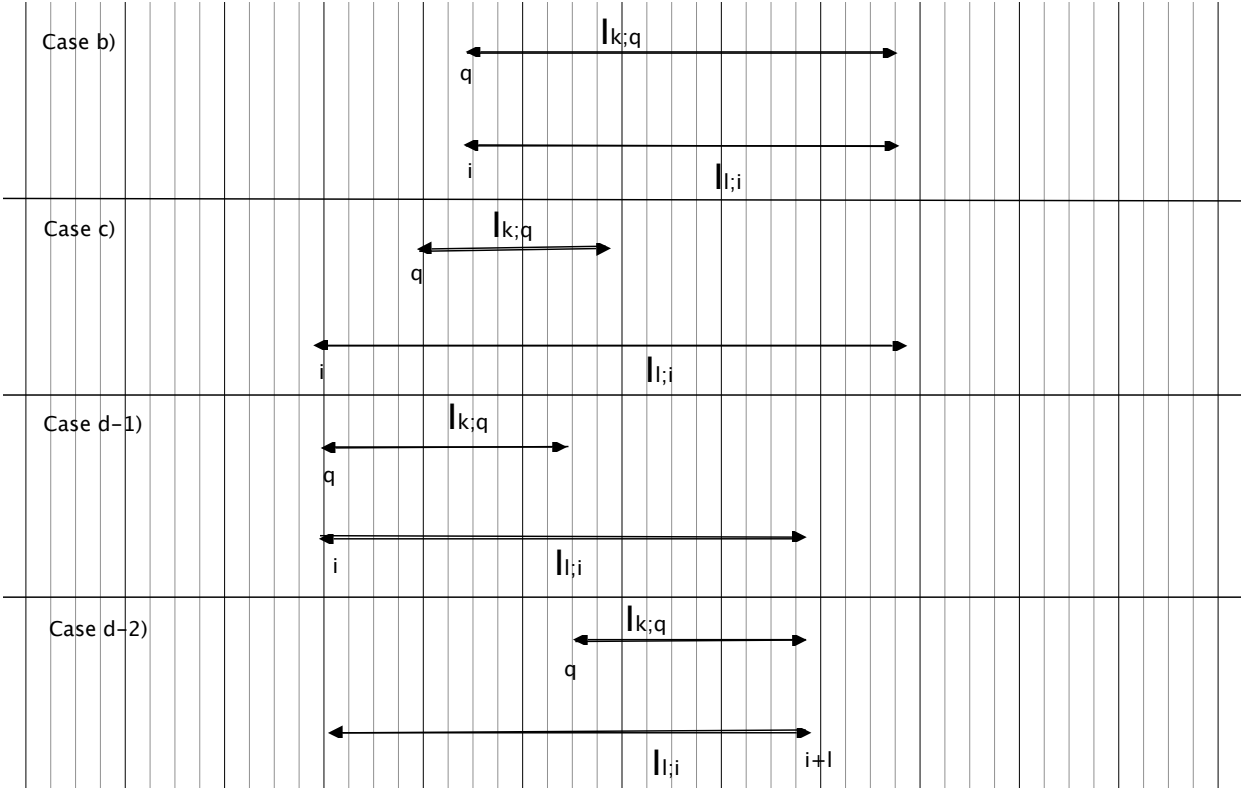}
 \caption{Relative positions of intervals $I_{k;q}$ and $I_{l;i}$}
 \label{fig:cases-bis}
\end{figure}

In this subsection, we finally present a precise \textit{iterative definition} of the operators $$V^{(k,q)}_{I_{l;i}}:=\alpha_{I_{k;q}}(V^{(k,q-1)}_{I_{l;i}})$$ in terms of the operators, $V^{(k,q-1)}_{I_{l;i}}$, at the previous step $(k,q-1)$, starting from 
\begin{equation}\label{initial-V}
V_{I_{0;i}}^{(0,N)}\equiv H_i\quad ,\quad
V_{I_{1;i}}^{(0,N)}\equiv V_{I_{1;i}}\quad,\quad
V_{I_{l;i}}^{(0,N)} =0\,\,\text{for}\,\, l\geq 2.
\end{equation}
\begin{defn}\label{def-interections}
We assume that, for fixed $(k,q-1)$, with $(k,q-1) \succ (0,N)$, the operators $V^{(k,q-1)}_{I_{l;i}}$ and $S_{I_{k;q}}$ are well defined and bounded, for any $l,i$; or  we assume that $(k,q)=(1,1)$ and that the operator $S_{I_{1;1}}$ is well defined. We then define the operators $V^{(k,q)}_{I_{l;j}}$ as follows, with the warning that  if $q=1$ the couple $(k,q-1)$ is replaced by $(k-1,N-k+1)$ in eqs. (\ref{case-in})-(\ref{main-def-V-bis}) -- see Fig. \ref{fig:cases-bis} for a graphical representation of the different cases b), c) d-1) and d-2, below:
\begin{itemize}
\item[a)]
in all the following cases
\begin{itemize}
\item[a-i)] $l\leq k-1$;
\item[a-ii)]  $I_{l;i}\cap I_{k;q}=\emptyset$;
\item[a-iii)]  $ I_{l;i}\cap I_{k;q}\neq\emptyset$ but $l\geq k$ and $I_{k;q} \nsubseteq I_{l;i}$;
\end{itemize}
we define
\begin{equation}\label{case-in}
V^{(k,q)}_{I_{l;i}}:=V^{(k,q-1)}_{I_{l;i}}\,;
\end{equation}
\item[b)]
if $I_{l;i}\equiv I_{k;q}$, we define
\begin{equation}
V^{(k,q)}_{I_{l;i}}:= \sum_{j=1}^{\infty}t^{j-1}(V^{(k,q-1)}_{I_{l;i}})^{diag}_j \,;
\end{equation}
\item[c)]
 if $I_{k;q}\subset I_{l;i}$ and $i, i+l \notin I_{k;q}$, we define
\begin{equation}
V^{(k,q)}_{I_{l;i}}:=e^{S_{I_{k;q}}}\,V^{(k,q-1)}_{I_{l;i}}\,e^{-S_{I_{k;q}}}\,;
\end{equation}

\item[d)]

\noindent
if $I_{k;q}\subset I_{l;i}$ and either $i$ or $i+l$ belongs to $ I_{k;q}$, we define
\begin{itemize}
\item[ d-1)] if $i$ belongs to $ I_{k;q}$, i.e., $q \equiv i$,  then
\begin{eqnarray}
V^{(k,q)}_{I_{l;i}} &:= & e^{S_{I_{k;q}}}\,V^{(k,q-1)}_{I_{l;i}}\,e^{-S_{I_{k;q}}}\,+\sum_{j=1}^{k}\sum_{n=1}^{\infty}\frac{1}{n!}\,ad^{n}S_{I_{k;i}}(V^{(k,q-1)}_{I_{l-j;i+j}})\,; \label{main-def-V}
\end{eqnarray}

\item[ d-2)] if $i+l$ belongs to $ I_{k;q}$, i.e., $q +k\equiv i+l$ that means $q\equiv i+l-k$,  then
\begin{eqnarray}
V^{(k,q)}_{I_{l;i}} &:= & e^{S_{I_{k;q}}}\,V^{(k,q-1)}_{I_{l;i}}\,e^{-S_{I_{k;q}}}\,+\sum_{j=1}^{k}\sum_{n=1}^{\infty}\frac{1}{n!}\,ad^{n}S_{I_{k;i+l-k}}(V^{(k,q-1)}_{I_{l-j;i}})\,. \label{main-def-V-bis}
\end{eqnarray}
\end{itemize}
Notice that  in both cases, d-1) and d-2), the elements of the sets $\{I_{l-j;i+j}\}_{j=1}^{k}$ and $\{I_{l-j;i}\}_{j=1}^{k}$, respectively,  are all the intervals, $\mathscr{I}$, such that $\mathscr{I} \cap I_{k;q} \neq \emptyset $,  $\mathscr{I} \nsubseteq I_{k;q}$,  $I_{k;q}\nsubseteq \mathscr{I} $, and   $\mathscr{I} \cup I_{k;q}\equiv I_{l;i}$.
\end{itemize}
\end{defn}
\begin{rem}
Notice that, according to  Definition \ref{def-interections}:
\begin{itemize}
\item if $(k', q')\succ (l,i)$  then
\begin{equation}
V^{(k',q')}_{I_{l;i}}=V^{(l,i)}_{I_{l;i}}\,,
\end{equation}
since the occurrences in cases b), c), d-1), and d-2) are excluded;
\item
 for $k\geq 1$ and all  allowed choices of $q$, 
\begin{equation}
V^{(k,q)}_{I_{0;i}}=H_i\,
\end{equation}
due to a-i).
\end{itemize}
\end{rem}
In the next theorem we prove that Definition \ref{def-interections} yields operators $V_{I_{l;j}}^{(k,q)}$ consistent with the expression of the Hamiltonian $K_{N}^{(k,q)}$ given in Eq. (\ref{def-transf-ham})-(\ref{def-transf-ham-bis}).
\begin{thm}\label{th-potentials}
The Hamiltonian $K_{N}^{(k,q)}:=e^{S_{I_{k;q}}}\,K_N^{(k,q-1)}\,e^{-S_{I_{k;q}}}$, with $k\geq1$ and $q\geq 2$, has the form given in (\ref{def-transf-ham})-(\ref{def-transf-ham-bis}),   where the operators $\{V^{(k,q)}_{I_{l;i}}\}$ are determined by the operators $\{V^{(k,q-1)}_{I_{l;i}}\}$ of the previous iteration step, as specified in  Definition \ref{def-interections}. If $q=1$ the statement holds with $(k,q-1)$ replaced by $(k-1, N-k+1)$.
\end{thm}

\noindent
\emph{Proof.}

\noindent
We study the case $q\geq 2$ explicitly, the case $q=1$ is proven in the same way. In the expression
\begin{eqnarray}
e^{S_{I_{k;q}}}\,K_N^{(k,q-1)}\,e^{-S_{I_{k;q}}}&=&
e^{S_{I_{k;q}}}\,\Big[\sum_{i=1}^{N}H_{i}+t\sum_{i=1}^{N-1}V^{(k,q-1)}_{I_{1;i}}+t\sum_{i=1}^{N-2}V^{(k,q-1)}_{I_{2;i}}+\dots+t\sum_{i=1}^{N-k}V^{(k,q-1)}_{I_{k;i}} \nonumber \\
&+ &\sum_{i=1}^{N-k-1}V^{(k,q-1)}_{I_{k+1;i}}+\dots+t\sum_{i=1}^{2}V^{(k,q-1)}_{I_{N-2;i}}+tV^{(k,q-1)}_{I_{N-1;1}}\Big]e^{-S_{I_{k;q}}}\,
\end{eqnarray}
we observe that:

\begin{itemize}
\item
For all intervals  $I_{l;i}$ with the property  that $I_{l;i} \cap I_{k;q}=\emptyset$, 
\begin{equation}
e^{S_{I_{k;q}}}V^{(k,q-1)}_{I_{l;i}}e^{-S_{I_{k;q}}}=V^{(k,q-1)}_{I_{l;i}}=:V^{(k,q)}_{I_{l;i}}
\end{equation}
which follows from a-ii), Definition \ref{def-interections}.
\item
With regard to the terms constituting $G_{I_{k;q}}$ (see definition (\ref{def-G})), we get, after adding $tV^{(k,q-1)}_{I_{k;q}}$, 

\quad $e^{S_{I_{k;q}}}\,(G_{I_{k;q}}+tV^{(k,q-1)}_{I_{k;q}})\,e^{-S_{I_{k;q}}}=$
\begin{eqnarray}
&= &\sum_{i\subset I_{k;q} }H_i+t\sum_{I_{1;i} \subset I_{k;q}} V^{(k, q-1)}_{I_{1;i}}+\dots+t\sum_{I_{k-1;i}\subset I_{k;q}}V^{(k, q-1)}_{I_{k-1;i}}
+t\sum_{j=1}^{\infty}t^{j-1}(V^{(k,q-1)}_{I_{k;q}})^{diag}_j \nonumber\\
&= &\sum_{i\subset I_{k;q} }H_i+t\sum_{I_{1;i} \subset I_{k;q}} V^{(k, q)}_{I_{1;i}}+\dots+t\sum_{I_{k-1;i}\subset I_{k;q}}V^{(k, q)}_{I_{k-1;i}}
+tV^{(k,q)}_{I_{k;q}}\,\,,
\end{eqnarray}
where the first identity is the result of the Lie-Schwinger conjugation and the last identity follows from Definition \ref{def-interections}, cases a-i) and b).
\item
With regard to the terms $V^{(k,q-1)}_{I_{l;i}}$,  with $ I_{k;q} \subset I_{l;i}$ and $i,i+l \notin  I_{k;q}$, the expression
\begin{equation}
e^{S_{I_{k;q}}}\,V^{(k,q-1)}_{I_{l;i}}e^{-S_{I_{k;q}}}
\end{equation}
corresponds to $V^{(k,q)}_{I_{l;i}}$, by Definition \ref{def-interections}, case c).
\item
With regard to the terms $V^{(k,q-1)}_{I_{l;i}}$, with  $I_{l;i}\cap I_{k;q}\neq\emptyset$, but $I_{l;i} \nsubseteq I_{k;q}$ and $I_{k;q} \nsubseteq I_{l;i}$, it follows that
\begin{equation}
e^{S_{I_{k;q}}}\,V^{(k,q-1)}_{I_{l;i}}\,e^{-S_{I_{k;q}}}=V^{(k,q-1)}_{I_{l;i}}+\sum_{n=1}^{\infty}\frac{1}{n!}\,ad^{n}S_{I_{k;q}}(V^{(k,q-1)}_{I_{l;i}}) \,.\label{growth}
\end{equation}
The first term on the right side is $V^{(k,q)}_{I_{l;i}}$ (see cases a-i) and a-iii) in Definition \ref{def-interections}), the second term contributes to $V^{(k,q)}_{I_{r;j}}$, where $I_{r;j} \equiv I_{l;i}\cup I_{k;q}$,  together with further similar terms and with
\begin{equation}\label{extra}
e^{S_{I_{k;q}}}\,V^{(k,q-1)}_{I_{r;j}}\,e^{-S_{I_{k;q}}}\,,
\end{equation}
where the set $I_{r;j}$ has the property that $I_{k;q}\subset I_{r;j}$, and either $j$ or $j+r$ belong to $ I_{k;q}$. Notice that the term in (\ref{extra})  has not been considered in the previous cases and corresponds to the first term in (\ref{main-def-V}) or in (\ref{main-def-V-bis}),  where $l$ is replaced by $r$ and $i$ by $j$. \qed
\end{itemize}

\subsection{Block-diagonalization of $K_{N}$ - control of $\|V^{(k,q)}_{I_{r;i}}\|$}
In the next theorem, we estimate the norm of $V^{(k,q)}_{I_{r;i}}$ in terms of the norm of $V^{(k,q-1)}_{I_{r;i}}$. For a fixed interval $I_{r;i}$, the norm of the potential does not change, i.e., $\|V^{(k,q-1)}_{I_{r;i}}\|=\|V^{(k,q)}_{I_{r;i}}\|$, 
in the step $(k,q-1) \rightarrow (k,q)$, unless some conditions are fulfilled. To gain some intuition of this fact, the reader is advised to take a look at Fig. 1, (replacing $l$ by $r$). Notice that shifting the interval $I_{k;q}$ to the left by one site makes it coincide with $I_{k;q-1}$. If $I_{k;q}$ is not contained in $I_{r;i}$ or if it is contained therein, but none of the endpoints  of the interval $I_{k,q}$  coincides with an  endpoint of $I_{r;i}$,  then  $\|V^{(k,q)}_{I_{r;i}}\|=\|V^{(k,q-1)}_{I_{r;i}}\|$. Therefore, if an interval of length $k$ is shifted ``to the left",  a change of norm, i.e., 
$\|V^{(k,q)}_{I_{r;i}}\|\neq \|V^{(k,q-1)}_{I_{r;i}}\|$, only happens in  at most two cases, provided $r> k$, and only in one case if $k$ coincides with the length $r$\,; and it never happens if $r<k$.  

\noindent
In the theorem below we estimate the change of the norm of the potentials in the block-diagonalization steps, for each $k$, starting from $k=0$. It is crucial to control the block-diagonalization of $V^{(k,q)}_{I_{r;i}}$ that takes place when 
$(k,q)\equiv (r,i)$. In this step,  we have to make use of a lower bound on the gap above the ground-state energy in the energy spectrum of the Hamiltonian $G_{I_{k;q}}$. This lower bound follows from estimate (\ref{ass-2}), as explained in Lemma \ref{gap} and Corollary \ref{cor-gap}. We will proceed inductively by showing that, for $t$ sufficiently small but independent of $N$, $k$, and $q$, the operator-norm bound in (\ref{ass-2}), at step $(k,q-1)$, $q\geq 2$  (for $q=1$ see the footnote), yields control over the spectral gap of the Hamiltonians $G_{I_{k;q}}$, (see Corollary \ref{cor-gap}), and the latter provides an essential  ingredient for the proof of a bound on the operator norms of the potentials, according to  (\ref{ass-2}), at the next step\footnote{  \label{footnote} Recall the special steps of type $(k,1)$ with preceding step $(k-1, N-k+1)$.} $(k,q)$.

\begin{thm}\label{th-norms}
Assume that the coupling constant $t>0$ is sufficiently small. Then the Hamiltonians $G_{I_{k;q}}$ and $K_{N}^{(k,q)}$ are well defined, and
\begin{enumerate}
\item[S1)]{ for any interval  $I_{r;i}$, with $r\geq 1$, the operator $V^{(k,q)}_{I_{r;i}}$ has a norm bounded by $\frac{8}{(r+1)^2}\,t^{\frac{r-1}{3}}$,}
\item[S2)]{ $G_{I_{k;q+1}}$ has a spectral gap $\Delta_{I_{k,q+1}}\geq \frac{1}{2} $ above the ground state energy,
where  $G_{I_{k;q}}$ is defined in (\ref{def-G}) for $k\geq 2$,  and $G_{I_{1;q}}:=H_{q}+H_{q+1}$.}
\end{enumerate}
\end{thm}

\noindent
\emph{Proof.}
The proof is by induction in the diagonalization step $(k,q)$, starting at $(k,q)=(0,N)$, and ending at $(k,q)=(N-1,1)$; (notice that \textit{S2)}  is not defined for $(k,q)=(N-1,1)$).  

\noindent
For $(k,q)= (0,N)$, we observe that $K^{(k,q)}_N\equiv K_N$ and $G_{I_{k;q}}$ is not defined, indeed it is not needed since   \textit{S1)} is verified by direct computation,  because by definition
$$\|V_{I_{1;i}}^{(0,N)}\|=\| V_{I_{1;i}}\|=1\,,$$
and
$V_{I_{r;i}}^{(0,N)} =0$, for $r\geq 2$. \textit{S2)} holds trivially since, by definition,  the successor of $(0,N)$ is $(1,1)$ and $G_{I_{1;1}}=H_1+H_2$.  

\noindent
Assume that \textit{S1)} and \textit{S2)} hold for all  steps $(k',q')$ with $(k',q') \prec (k,q)$. We prove that they then hold at step $(k,q)$. By Lemma \ref{control-LS}, \textit{S1)} and \textit{S2)} for $ (k,q-1)$ imply that $S_{I_{k;q}}$ and, consequently, that $K_{N}^{(k,q)}$ are well defined operators, (see (\ref{def-K})). In the steps described  below it is understood that if $q=1$ the couple $(k,q-1)$ is replaced by $(k-1,N-k+1)$.
\\

\noindent
\emph{Induction step in the proof of S1)}

\noindent
Starting from Definition  \ref{def-interections} we consider the following cases:

\noindent
\emph{Case $r=1$.}

\noindent
Let $k>1(=r)$ or $k=1=r$ but $I_{1;i}$ such that $i \neq q$.  Then the possible cases are described in a-i), a-ii),  and a-iii), see Definition  \ref{def-interections}, and we have that
\begin{equation}
\|V^{(k,q)}_{I_{1;i}}\|=\|V^{(k,q-1)}_{I_{1;i}}\|\,.
\end{equation} 

\noindent
Hence, we use the inductive hypothesis.

\noindent
Let $k=1$ and  assume the set $I_{1;i}$ is equal to $I_{1;q}$. Then  we refer to case b) and we  find that
\begin{equation}
\|V^{(1,q)}_{I_{1;q}}\| \leq 2 \|V^{(1,q-1)}_{I_{1;q}}\|=2\,,
\end{equation}
where:
\begin{itemize}
\item[a)] the inequality  $\|V^{(1,q)}_{I_{1;q}}\| \leq 2 \|V^{(1,q-1)}_{I_{1;q}}\|$ holds for $t$ sufficiently small uniformly in $q$ and $N$,  thanks to  Lemma \ref{control-LS} which can be applied since we assume S1) and S2) at step $(1,q-1)$;
\item[b)]  we use $ \|V^{(1,q-1)}_{I_{1;q}}\|=\|V^{(1,q-2)}_{I_{1;q}}\|=\dots =\|V^{(0,N)}_{I_{1;q}}\|=1$. 
\end{itemize}

\noindent
\emph{Case $r\geq 2$.}

\noindent
{\bf{I)}} 

\noindent
Let $r>k$, then in cases a-ii), a-iii), and c),  see Definition  \ref{def-interections}, we have that
\begin{equation}\label{cons}
\|V^{(k,q)}_{I_{r;i}}\|=\|V^{(k,q-1)}_{I_{r;i}}\|\,,
\end{equation}  
otherwise we are in  case d-1), see  (\ref{main-def-V}), that means $q\equiv i$, and estimate
\begin{equation}\label{arg}
\|V^{(k,q)}_{I_{r;i}}\|\leq \|V^{(k,q-1)}_{I_{r;i}}\|+\sum_{j=1}^{k}\sum_{n=1}^{\infty}\frac{1}{n!}\,\| ad^{n}S_{I_{k;i}}(V^{(k,q-1)}_{I_{r-j;i+j}})\|\,,
\end{equation}
or in case d-2), see  (\ref{main-def-V-bis}), that means $q\equiv i+r-k$,  and we estimate
\begin{equation}\label{arg-bis}
\|V^{(k,q)}_{I_{r;i}}\|\leq \|V^{(k,q-1)}_{I_{r;i}}\|+\sum_{j=1}^{k}\sum_{n=1}^{\infty}\frac{1}{n!}\,\| ad^{n}S_{I_{k;i+r-k}}(V^{(k,q-1)}_{I_{r-j;i}})\|\,.
\end{equation}
Hence, for $r>k$ and $q> i+r-k$, we can write (assuming $i\geq 2$, but an analogous procedure holds if $i=1$) 
\begin{eqnarray}
& &\|V^{(k,q)}_{I_{r;i}}\|\label{a} \\
& =&\|V^{(k,i+r-k)}_{I_{r;i}}\|\label{b} \\
&\leq &\|V^{(k,i+r-k-1)}_{I_{r;i}}\|+\sum_{j=1}^{k}\sum_{n=1}^{\infty}\frac{1}{n!}\,\| ad^{n}S_{I_{k;i+r-k}}(V^{(k,i+r-k-1)}_{I_{r-j;i}})\| \label{c} \,\\
&=&\|V^{(k,i)}_{I_{r;i}}\|+\sum_{j=1}^{k}\sum_{n=1}^{\infty}\frac{1}{n!}\,\| ad^{n}S_{I_{k;i+r-k}}(V^{(k,i+r-k-1)}_{I_{r-j;i}})\| \label{d} \\
&\leq &\|V^{(k,i-1)}_{I_{r;i}}\|+\sum_{j=1}^{k}\sum_{n=1}^{\infty}\frac{1}{n!}\,\| ad^{n}S_{I_{k;i}}(V^{(k,i-1)}_{I_{r-j;i+j}})\|+\sum_{j=1}^{k}\sum_{n=1}^{\infty}\frac{1}{n!}\,\| ad^{n}S_{I_{k;i+r-k}}(V^{(k,i+r-k-1)}_{I_{r-j;i}})\|\quad \label{e} \\
&=&\|V^{(k-1,i+r-k+1)}_{I_{r;i}}\|+\sum_{j=1}^{k}\sum_{n=1}^{\infty}\frac{1}{n!}\,\| ad^{n}S_{I_{k;i}}(V^{(k,i-1)}_{I_{r-j;i+j}})\|+\sum_{j=1}^{k}\sum_{n=1}^{\infty}\frac{1}{n!}\,\| ad^{n}S_{I_{k;i+r-k}}(V^{(k,i+r-k-1)}_{I_{r-j;i}})\|\label{f} \quad\quad\quad
\end{eqnarray}
where 
\begin{itemize}
\item in the step from (\ref{a}) to (\ref{b}) we have used that $\|V^{(k,q')}_{I_{r;i}}\|=\|V^{(k,q'-1)}_{I_{r;i}}\|$, for all $q\geq q'\geq i+r-k+1$, by the argument yielding (\ref{cons});
\item in the step from (\ref{b}) to (\ref{c}) we have used (\ref{arg-bis});
\item in the step from (\ref{c}) to (\ref{d})  we have used that $\|V^{(k,q')}_{I_{r;i}}\|=\|V^{(k,q'-1)}_{I_{r;i}}\|$, for all $i+r-k-1\geq q'\geq i+1$, by the argument yielding (\ref{cons}), (notice that if $r=k+1$ this step is not needed);
\item in the step from (\ref{d}) to (\ref{e}) we have used (\ref{arg});
\item in the step from (\ref{e}) to (\ref{f}),   invoking the argument yielding (\ref{cons}),  we have used the following facts: 
\begin{enumerate} 
\item[a)] $\|V^{(k,q')}_{I_{r;i}}\|=\|V^{(k,q'-1)}_{I_{r;i}}\|$, for all $i-1\geq q'\geq 2$,
\item[b)]  $\|V^{(k,1)}_{I_{r;i}}\|=\|V^{(k-1,N-k+1)}_{I_{r;i}}\|$,
\item[c)]   $\|V^{(k-1,q')}_{I_{r;i}}\|=\|V^{(k-1,q'-1)}_{I_{r;i}}\|$, for all $N-k+1\geq q'\geq i+r-k+2$.
\end{enumerate}
Notice that if $i=2$  then a) is an empty statement and must be ignored, likewise if $N=i+r$  statement c) is empty and must be ignored.
\end{itemize}

\begin{figure}
 \includegraphics[width=\linewidth]{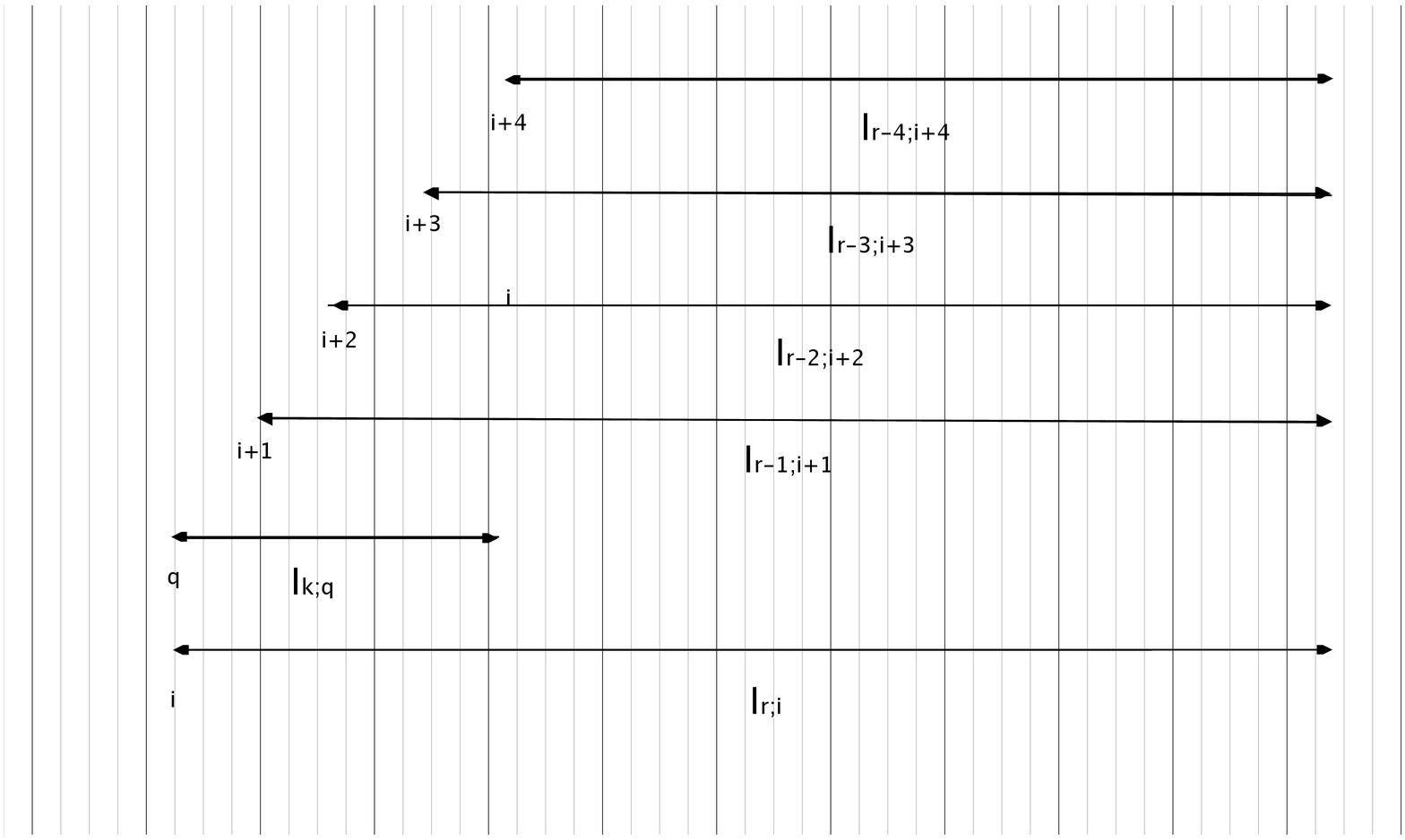}
 \caption{Intervals $I_{k;q}$ and $I_{r-j;i+j}$, $j=1,2,3,4$, associated with the summands in formula (\ref{blocks}) for $m=k$ and $r=l$.}
 \label{fig-3}
\end{figure}

Iterating the arguments in (\ref{cons}) and (\ref{arg})-(\ref{arg-bis})  for the first term on the right side of (\ref{f}) -- i.e.,  $\|V^{(k-1,i+r-k+1)}_{I_{r;i}}\|$) -- and observing that, by assumption, $V^{(0,N)}_{I_{r;i}}=0$ if $r\geq 2$, we end up finding that
\begin{eqnarray}
\|V^{(k,q)}_{I_{r;i}}\| &\leq &\sum_{m=1}^{k}\sum_{j=1}^{m}\sum_{n=1}^{\infty}\frac{1}{n!}\,\|ad^{n}S_{I_{m; i+r-m}}(V^{(m, i+r-m-1)}_{I_{r-j;i}}) \| \label{blocks-0}\\
& &+\sum_{m=1}^{k}\sum_{j=1}^{m}\sum_{n=1}^{\infty}\frac{1}{n!}\,\|ad^{n}S_{I_{m;i}}(V^{(m,i-1)}_{I_{r-j;i+j}})\|\,, \label{blocks}
\end{eqnarray}
where the interval $I_{m; i}$, with $m\leq k-1$, has first vertex coinciding with $i$ and last vertex equal to $i+m$, while  the set $I_{m; i+r-m}$  has first vertex equal to $i+r-m$ and the last vertex coinciding with $i+r$. The intervals associated to the summands in (\ref{blocks}) are displayed in Fig. \ref{fig-3}. Notice that the last two terms on the right side of (\ref{f}) correspond to the summands associated with $m \equiv k$ in (\ref{blocks-0}) and (\ref{blocks}).
\\

\noindent
{\bf{II)}} 

\noindent
Let $r=k$, then the interval is $I_{r \equiv k;i}$. 

\noindent
If  $i>q $ then the procedure is identical to the previous case {\bf{I)}} except that the sum over $m$ in (\ref{blocks-0})-(\ref{blocks}) is up to $k-1$. 

\noindent
If $i\leq q$ we have two possibilities:
\begin{itemize}
\item[a)] if $i<q$ then $\|V^{(k,q)}_{I_{r\equiv k;i}}\| =\|V^{(k,q-1)}_{I_{r\equiv k;i}}\|$;
\item[b)]  if  $q = i$ we refer to case b) (of Definition  \ref{def-interections}) and, thanks to S1) and S2) of the previous step $(k,q-1)$, we can apply Lemma \ref{control-LS} and estimate
\begin{equation}
\|V^{(k,q\equiv i)}_{I_{r\equiv k;i}}\| \leq 2 \|V^{(k,i-1)}_{I_{r\equiv k;i}}\|\,.
\end{equation}
Then we proceed as in the previous case {\bf{I)}}. Eventually we can estimate
\begin{eqnarray}\label{general-estimate}
\|V^{(k,q)}_{I_{r\equiv k;i}}\| &\leq & 2 \sum_{m=1}^{k-1}\sum_{j=1}^{m}\sum_{n=1}^{\infty}\frac{1}{n!}\,\|ad^{n}S_{I_{m; i+r-m}}(V^{(m, i+r-m-1)}_{I_{r-j;i}}) \| \\
& &+2 \sum_{m=1}^{k-1}\sum_{j=1}^{m}\sum_{n=1}^{\infty}\frac{1}{n!}\,\|ad^{n}S_{I_{m;i}}(V^{(m,i-1)}_{I_{r-j;i+j}})\| \label{formula-fig}
\label{general-estimate-bis}
\end{eqnarray}
\end{itemize}

\noindent
{\bf{III)}} 

\noindent
Let $r<k$. This corresponds to case a-i) in Definition  (\ref{def-interections}). Thus 
\begin{equation}
\|V^{(k,q)}_{I_{r;i}}\|=\|V^{(k,q-1)}_{I_{r;i}}\|\,,
\end{equation} 

\noindent
and we use the inductive hypothesis.
\\

\noindent
\emph{Estimate of (\ref{general-estimate})-(\ref{general-estimate-bis})}

\noindent
We recall formula (\ref{formula-S}), and we invoke properties S1) and S2) at the previous steps, so that we can apply Lemma \ref{control-LS} and get the estimate
\begin{equation}
\|S_{I_{m; i+r-m}}\|\leq C\cdot t \cdot \| V^{(m, i+r-m-1)}_{I_{m;i+r-m}}\|\leq C\cdot t \cdot \frac{8}{(m+1)^2}\,t^{\frac{m-1}{3}}\, \label{est-S}
\end{equation}
and
\begin{equation}
\|\sum_{n=1}^{\infty}\frac{1}{n!}\,\|ad^{n}S_{I_{m; i+r-m}}(V^{(m, i+r-m-1)}_{I_{r-j;i}})\|\leq C\cdot t \cdot \frac{64}{(m+1)^2}\,t^{\frac{m-1}{3}}\frac{t^{\frac{r-j-1}{3}}}{(r-j+1)^2}\,, \label{est-series}
\end{equation}
where the value of the constant $C$ changes from line to line.  

\noindent
Using the observations on the construction of the intervals (see (\ref{blocks})), the estimate in (\ref{est-series}), and property S1)  at the previous steps, we can write
\begin{eqnarray}
\|V^{(k,q)}_{I_{r;i}}\|&\leq &128\cdot C\cdot t\, \sum_{m=1}^{r-1}\frac{t^{\frac{m-1}{3}}}{(m+1)^2}\,\sum_{j=1}^{m}\,\frac{t^{\frac{r-j-1}{3}}}{(r-j+1)^2}\nonumber\\
&\leq &C'\cdot t\,\sum_{m=1}^{r-1}\frac{t^{\frac{m-1}{3}}}{m^2}\,\frac{t^{\frac{r-m-1}{3}}}{(r-m)^2}\nonumber\\
&\leq &C'\cdot t\,\cdot t^{\frac{r-2}{3}}\sum_{m=1}^{r-1}\frac{1}{m^2}\,\frac{1}{(r-m)^2}\nonumber\\
&\leq &\frac{8\cdot t^{\frac{r-1}{3}}}{(r+1)^2}\,,
\end{eqnarray}
for $t>0$ sufficiently small, but independent of $r\geq 2$, $k$, and $q$, where $C'$ is a universal constant.

The estimate of (\ref{blocks-0})-(\ref{blocks}) is similar.
\\

\noindent
\emph{Induction step to prove S2)}

\noindent
Having proven \textit{S1)}, we can use Lemma \ref{gap} and Corollary \ref{cor-gap} in subsequent arguments. Hence, \textit{S2)} holds for $t$ sufficiently small, but independent of $N$, $k$, and $q$. \qed
\setcounter{equation}{0}

\begin{thm}\label{main-res}
Under the assumption that (\ref{gaps}), (\ref{potential}) and (\ref{norms}) hold, the Hamiltonian $K_{N}$ defined in (\ref{Hamiltonian}) has the following properties: There exists some $t_0 > 0$ such that, for any $t\in \mathbb{R}$ with 
$\vert t \vert < t_0$, and for all $N < \infty$,
\begin{enumerate}
\item[(i)]{ $K_{N}\equiv K_N(t)$ has a unique ground-state; and}
\item[(ii)]{ the energy spectrum of $K_N$ has a strictly positive gap, $\Delta_{N}(t) \geq \frac{1}{2}$, above the ground-state energy.}
\end{enumerate}
\end{thm}

\noindent
\emph{Proof.}
Notice that $K_N^{(N-1,1)} \equiv G_{I_{N-1;1}}+tV^{(N-1,1)}_{I_{N-1;1}}$. We have constructed the unitary conjugation 
$\text{exp}S_{N}(t)$,  (see eq. (\ref{conjug})), such that  the operator
$$e^{S_{N}(t)}K_{N}(t)e^{-S_{N}(t)}=G_{I_{N-1;1}}+tV^{(N-1,1)}_{I_{N-1;1}}=: \widetilde{K}_{N}(t),$$  
has the properties in Eqs. (\ref{block-diag}) and (\ref{gapss}), which follow from Theorem \ref{th-norms} and from Eqs. (\ref{final-eq-1}) and (\ref{final-eq-2}), for $(k,q)=(N-1,1)$, where we also include the block-diagonalized potential $V^{(N-1,1)}_{I_{N-1;1}}$. \qed

\section{The Kitaev chain}
In this last section, we show how our method can be used to study small perturbations of the Hamiltonian of a Kitaev chain in the nontrivial phase; (see \cite{GST}). (Similar results have been proven in \cite{KST} for a special class of perturbations.)

Consider a chain with $N$ sites, where, at each site $j$, there are fermion creation- and annihilation operators, $c^{\dagger}_j,\, c_j$, with 
\begin{equation}\label{anti-rel}
\{c_j\,,\,c_{l}\}=\{c^{\dagger}_j\,,\,c^{\dagger}_{l}\}=0 \quad,\quad \{c_j\,,\,c^{\dagger}_l\}=\delta_{j,l}\,,
\end{equation}
where $\lbrace A, B \rbrace$ is the anti-commutator of $A$ and $B$. The Hilbert space $ \mathcal{H}$ is spanned by the vectors obtained by applying products of creation operators, $c^{\dagger}_j$, $j=1,\dots,N$, to the vacuum vector, which is annihilated by all the operators $c_j$.
The Hamiltonian of the system  is given by
\begin{equation}
H:=-\mu \sum_{j=1}^{N}c^{\dagger}_jc_j-\sum_{j=1}^{N-1}\,\Big(\tau \, c_j^{\dagger}c_{j+1}+\tau \, c^{\dagger}_{j+1}c_j+\Delta c_jc_{j+1}+\Delta c^{\dagger}_{j+1}c^{\dagger}_j)
\end{equation}
where $\mu$ is the chemical potential, $\tau \geq 0$ is the nearest-neighbor hopping amplitude, and $\Delta\geq 0$ is the p-wave pairing amplitude. By re-writing the fermion operators $c_j$, $c_j^{\dagger}$ in terms of the Clifford  generators (``Dirac matrices'')
\begin{equation}
\gamma_{A,j}:=-ic_j^{\dagger}-ic_j\quad, \quad \gamma_{B,j}:=c_j^{\dagger}+c_j\,,
\end{equation}
the Hamiltonian becomes
\begin{equation}
H=-\frac{\mu}{2} \sum_{j=1}^{N}(1+i\gamma_{B,j}\gamma_{A,j})-\frac{ i}{2}\sum_{j=1}^{N-1}\,\{(\Delta +\tau )\gamma_{B,j}\gamma_{A,j+1} +(\Delta-\tau)\gamma_{B,j}\gamma_{A,j+1}\}\,.
\end{equation}
If $\mu=0$ and $\tau =\Delta=1$, and for open boundary conditions,  the system is in a ``nontrivial phase",  and the corresponding Hamiltonian is denoted by $H_{Kitaev}$:
\begin{equation}
H_{Kitaev}=-i\sum_{j=1}^{N-1}\,\gamma_{B,j}\gamma_{A,j+1} =\sum_{j=1}^{N-1}\,(2d^{\dagger}_j\,d_j-1)\,,
\end{equation}
where 
\begin{equation}
2d^{\dagger}_j:=\gamma_{B,j}+i\gamma_{A,j+1} =c_{j+1}-c^{\dagger}_{j+1}+c^{\dagger}_j+c_j\quad, \quad 2d^{\dagger}_0:=-c^{\dagger}_{1}+c_{1}+c^{\dagger}_N+c_N.
\end{equation}
As a consequence of (\ref{anti-rel}), the variables $d_j\,,d^{\dagger}_{j'}$ obey the relations
\begin{equation}
\{d_j\,,\,d_{l}\}=\{d^{\dagger}_j\,,\,d^{\dagger}_{l}\}=0 \quad,\quad \{d_j\,,\,d^{\dagger}_l\}=\delta_{j,l}\,\quad \text{for}\,\, \, \, j,l=0,\dots, N-1\,.
\end{equation}

\noindent
Notice that, for $1\leq j \leq N-1$,
\begin{equation}\label{invers-1}
c_j=\frac{d_j+d^{\dagger}_{j}+d^{\dagger}_{j-1}-d_{j-1}}{2}\,,
\end{equation}
and
\begin{equation}\label{invers-2}
c_N=\frac{d_{0}+d^{\dagger}_0+d^{\dagger}_{N-1}-d_{N-1}}{2}\,.
\end{equation}

Consider the following local perturbations of the Hamiltonian $H$:
\begin{equation}\label{local-1}
\beta \sum_{i=1}^{N-1}V_{I_{1;i}}+\beta \sum_{i=1}^{N-2}V_{I_{2;i}}+\dots +\beta\sum_{i=1}^{N-\bar{k}}V_{I_{\bar{k};i}}
\end{equation}
where $\bar{k}$ is $N$-independent,  $\beta>0$ is a coupling constant,  and  each term
\begin{equation}\label{local-op}
V_{I_{j;i}}
\end{equation}
is a hermitian operator consisting of an $N$-independent,  finite sum of products of an even number of operators $\{c_l\,,\,c^{\dagger}_{l}\}_{l=i}^{i+j}$. 
In (\ref{local-1}) we split the sum into
\begin{eqnarray}
\beta V_{I_{1;1}}&+&\beta V_{I_{2;1}}+\dots +\beta V_{I_{\bar{k};1}} \label{left}\\
&+&\beta \sum_{i=2}^{N-2}V_{I_{1;i}}+\beta \sum_{i=2}^{N-3}V_{I_{2;i}}+\dots +\beta\sum_{i=2}^{N-\bar{k}-1}V_{I_{\bar{k};i}}\label{center}\\
& &\,\,+ \beta V_{I_{1;N-1}}+\beta V_{I_{2;N-2}}+\dots +\beta V_{I_{\bar{k};N-\bar{k}}}\,,\label{right}
\end{eqnarray}
and we then use the identities (\ref{invers-1})-(\ref{invers-2}) to re-write these operators in terms of the $d$- and $d^{\dagger}$- variables. We then get
\begin{eqnarray}
(\ref{left})\,\,+\,\,(\ref{center})&+&(\ref{right})\\
&= &\beta \tilde{V}_{I_{2;0}}+\beta \tilde{V}_{I_{3;0}}+\dots +\beta \tilde{V}_{I_{\bar{k};0}} \\
&+&\beta \sum_{i=2}^{N-2}\tilde{V}_{I_{2;i-1}}+\beta \sum_{i=2}^{N-3}\tilde{V}_{I_{3;i-1}}+\dots +\beta\sum_{i=2}^{N-\bar{k}-1}\tilde{V}_{I_{\bar{k}+1;i-1}}\label{middle}\\
&+&\beta \tilde{V}_{I_{2;N-2}}+\beta \tilde{V}_{I_{3;N-3}}+\dots +\beta \tilde{V}_{I_{\bar{k}+1;N-\bar{k}-1}}\,,\label{last-eq}
\end{eqnarray}
where, in (\ref{last-eq}), $N$ is identified with $0$, i.e, $I_{2;N-2}:=(N-2,N-1,0)$, and  the symbol
\begin{equation}
\tilde{V}_{I_{j;i}}
\end{equation}
stands for a finite sum of operators consisting of products of an even number of operators $\{d_l\,,\,d^{\dagger}_{l}\}_{l=i}^{i+j}$.

Let $\Omega^{(d)}$ be the vector annihilated by the operators $d_j$, $j=0,\dots, N-1$, and define the Fock space 
$\mathcal{F}_{1,\dots,N-1}$ as the span of vectors obtained by applying products of the operators $d^{\dagger}_j$, $j=1,\dots, N-1$, to $\Omega^{(d)}$. This space can be identified with the space
\begin{equation}
\bigotimes_{j=1}^{N-1}\mathcal{F}_j\,,
\end{equation}
where $\mathcal{F}_j \simeq \mathbb{C}^{2}$ is the fermionic Fock space obtained by applying the identity and the creation operator $d^{\dagger}_j$ to the vacuum vector $\Omega^{(d)}_j$, which is annihilated by $d_j$. Likewise, we have that
\begin{equation}
\mathcal{H}\simeq \mathcal{F}_0\otimes \Big(\bigotimes_{j=1}^{N-1}\mathcal{F}_j\Big)\,.
\end{equation}
\noindent
Notice that the operators in (\ref{middle}) do not depend on the zero-mode operator. Hence we can apply the method developed in previous sections to analyse the Hamiltonian
\begin{equation}\label{intermediate}
H'_{\beta} \upharpoonright \,\bigotimes_{j=1}^{N-1}  \mathcal{F}_j:=\Big(H_{Kitaev}+\beta \sum_{i=2}^{N-2}\tilde{V}_{I_{2;i-1}}+\beta \sum_{i=2}^{N-3}\tilde{V}_{I_{3;i-1}}+\dots +\beta\sum_{i=2}^{N-\bar{k}-1}\tilde{V}_{I_{\bar{k}+1;i-1}}\Big) \upharpoonright \,\bigotimes_{j=1}^{N-1}  \mathcal{F}_j
\end{equation}
and show that, for $\beta$ sufficiently small, there is a  unique ground-state and the energy spectrum is gapped, with a gap larger than $1$ above the ground-state energy, uniformly in $N$.

It is straightforward to check that the Hamiltonian $H'_{\beta}\,:\, \mathcal{H} \rightarrow \mathcal{H}$  has the same spectrum as $H'_{\beta} \upharpoonright \,\bigotimes_{j=1}^{N-1}  \mathcal{F}_j$\,; but, for each eigenvalue $E$, the corresponding eigenspace  is doubled, since if $\Psi_E\in \bigotimes_{j=1}^{N-1}  \mathcal{F}_j$ is an eigenvector of (\ref{intermediate}) corresponding to the eigenvalue $E$ then both vectors, $\Omega_0\otimes \Psi_E$ and $d^{\dagger}_0\Omega_0\otimes \Psi_E$, are eigenvectors corresponding to the \textit{same} eigenvalue $E$ of the operator $H'_{\beta}\,:\, \mathcal{H} \rightarrow \mathcal{H}$.  We denote by $\mathcal{H}_{\beta,gs}$ the doubly-degenerate ground-state subspace of $H'_{\beta}\,:\, \mathcal{H} \rightarrow \mathcal{H}$.

We can now apply our Lie-Schwinger block-diagonalization procedure to the operator
\begin{eqnarray}
H_{\beta}:&=&H'_{\beta}\\
& &+\beta \tilde{V}_{I_{2;0}}+\beta \tilde{V}_{I_{3;0}}+\dots +\beta \tilde{V}_{I_{\bar{k};0}} \\
& &+\beta \tilde{V}_{I_{2;N-2}}+\beta \tilde{V}_{I_{3;N-3}}+\dots +\beta \tilde{V}_{I_{\bar{k}+1;N-\bar{k}-1}}
\end{eqnarray}
by considering $H'_{\beta}$ as the unperturbed Hamiltonian: By constructing a unitary operator $U$ (as explained in the previous section), we can block-diagonalise $H_{\beta}$, so that the transformed Hamiltonian $U^{*}H_{\beta}U$ has the property
\begin{equation}
U^{*}H_{\beta}U\,:\, \mathcal{H}_{\beta,gs} \rightarrow \mathcal{H}_{\beta,gs}\quad, \quad U^{*}H_{\beta}U\,:\,(\mathcal{H}\ominus \mathcal{H}_{\beta,gs}) \rightarrow (\mathcal{H}\ominus \mathcal{H}_{\beta,gs})\,.
\end{equation}
The distance between the spectrum of $U^{*}H_{\beta}U \upharpoonright \mathcal{H}_{\beta,gs}$ and  the one of~$U^{*}H_{\beta}U \upharpoonright (\mathcal{H}\ominus \mathcal{H}_{\beta,gs})$ is of order $1$ provided $\beta$ is sufficiently small. Moreover, the operator $U^{*}H_{\beta}U \upharpoonright \mathcal{H}_{\beta,gs}$ is a \mbox{$2\times 2$ matrix} that can be diagonalised.
\qed

\setcounter{equation}{0}
\begin{appendix}
\section{Appendix}
\begin{lem}\label{op-ineq-1} For any $1\leq n \leq N$
\begin{equation}\label{main-ineq}
\sum_{i=1}^{n} P^{\perp}_{\Omega_i} \geq \charf -\bigotimes_{i=1}^{n} P_{\Omega_i}=:\,\Big(\bigotimes_{i=1}^{n} P_{\Omega_i}\Big)^{\perp}
\end{equation}
where $P^{\perp}_{\Omega_i}=\charf-P_{\Omega_i}$.

\end{lem}

\noindent
\emph{Proof}

\noindent
We call $P_{vac}:=\bigotimes_{i=1}^{n} P_{\Omega_i}$ acting on $\mathcal{H}^{(n)}:=\bigotimes_{i=1}^{n} \mathcal{H}_i $.  We define
\begin{equation}
A_n:=\sum_{j=1}^{n}P_{\Omega_j}^{\perp}+P_{vac}\,.
\end{equation}
Notice that all operators $P_{\Omega_j}^{\perp}$ and $P_{vac}$ commute each other and are orthogonal projections.  Therefore we deduce that 
\begin{equation}\label{spec}
\text{spec}(A_n)\subseteq \{0,1,2,\dots, n+1\}\,.
\end{equation}
We will show that
\begin{equation}\label{range}
\text{Range}\,A_n=\mathcal{H}^{(n)}\,.
\end{equation}
If (\ref{range}) holds then $0\notin \text{spec}(A_n)$.  By (\ref{spec}) it then follows that 
\begin{equation}
A_n\geq \charf\,.
\end{equation}
Thus, we are left proving (\ref{range}).

\begin{itemize}
\item[(i)] Assume that $\psi$ is perpendicular to the range of $A_{n}$, and let
$P_{\Omega_j}^{\perp}\psi =:\phi_j$. Then, since $\psi \perp \text{Range}\, A_n$, we have that
\begin{equation}
0=\langle \psi, A_n \psi\rangle =\sum_{i\neq j}\langle \psi\,,\, P_{\Omega_i}^{\perp}\psi \rangle +\langle \psi\,\, P_{vac}\psi \rangle+\langle \psi\,,\, P_{\Omega_j}^{\perp}\psi\rangle\geq \langle \psi\,,\, P_{\Omega_j}^{\perp}\psi\rangle
\end{equation}
but 
\begin{equation}
\langle \psi\,,\, P_{\Omega_j}^{\perp}\psi\rangle =\langle P_{\Omega_j}^{\perp}\psi\,,\, P_{\Omega_j}^{\perp}\psi\rangle=\langle \phi_j\,,\,\phi_j\rangle
\end{equation}
where we have used that $P_{\Omega_j}^{\perp}$ is an orthogonal projection. We conclude that $\phi_j=0$ for all $j$.
\item[(ii)]
Let $\psi \perp \text{Range}\,A_n$. Then, by (i), 
\begin{equation}
\psi=\Big(\bigotimes_{j=1}^{n}\,(P_{\Omega_j}^{\perp}+P_{\Omega_j})\Big)\,\psi= (\bigotimes_{j=1}^{n}P_{\Omega_j})\,\psi=P_{vac}\psi 
\end{equation}
and 
\begin{equation}
0=\langle \psi\,,\, A_n\,\psi\rangle =\langle \psi\,,\, P_{vac}\,\psi\rangle =\langle \psi\,,\,\psi \rangle\quad \Rightarrow \quad \psi=0\,.
\end{equation}
\end{itemize}
Thus, $\text{Range}\, A_n=\mathcal{H}^{(n)}$, and (\ref{range}) is proven\,. \qed

From Lemma \ref{op-ineq-1} we derive the following bound.
\begin{cor}\label{op-ineq-2}
For $i+r\leq N$, we define
\begin{equation}
P^{(+)}_{I_{r;i}}:=\Big(\bigotimes_{k=i}^{i+r}P_{\Omega_{k}}\Big)^{\perp}\,.
\end{equation}
Then, for $1\leq l \leq L \leq N-r$, 
\begin{equation}\label{ineq-inter}
\sum_{i=l}^{L}P^{(+)}_{I_{r;i}}\leq (r+1) \sum_{i=l}^{L+r} P^{\perp}_{\Omega_i} \,.
\end{equation}
\end{cor}

\noindent
\emph{Proof}

\noindent
From Lemma \ref{op-ineq-1} we derive
\begin{equation}\label{first-lemma1.3}
\sum_{j=i}^{i+r} P^{\perp}_{\Omega_{j}}\geq \Big(\bigotimes_{k=i}^{i+r}P_{\Omega_{k}}\Big)^{\perp}\,.
\end{equation}
 By summing the l-h-s of (\ref{first-lemma1.3}) for $i$ from $l$ up to $L$, for each $j$ we get not more than $r+1$ terms of the type
$P^{\perp}_{\Omega_{j}}$
and the inequality in (\ref{ineq-inter}) follows\,. \qed

\begin{lem}\label{control-LS}
Assume that $t>0$ is sufficiently small,  $\|V^{(k,q-1)}_{I_{r;i}}\| \leq \frac{8}{(r+1)^2}\,t^{\frac{r-1}{3}}$, and $\Delta_{I_{k;q}}\geq \frac{1}{2}$. Then, for arbitrary $N$, $k\geq 1$, and $q\geq 2$, the inequalities
\begin{equation}\label{bound-V}
\|V^{(k,q)}_{I_{k;q}}\|\leq 2\|V^{(k,q-1)}_{I_{k;q}}\|\,
\end{equation}
\begin{equation}\label{bound-S}
\|S_{I_{k; q}}\|\leq C\cdot t \cdot \| V^{(k,q-1)}_{I_{k;q}}\|
\end{equation}
hold true for a universal constant $C$. For $q=1$,   $ V^{(k,q-1)}_{I_{k;q}}$  is replaced by $V^{(k-1,N-k+1)}_{I_{k;q}}$ in the right side of (\ref{bound-V}) and (\ref{bound-S}).
\end{lem}

\noindent
\emph{Proof.}

\noindent
In the following we assume $q\geq 2$; if $q=1$ an analogous proof holds. We recall that 
\begin{equation}
V^{(k,q)}_{I_{k;q}}:= \sum_{j=1}^{\infty}t^{j-1}(V^{(k,q-1)}_{I_{k;q}})^{diag}_j \,
\end{equation}
and
\begin{equation}
S_{I_{k;q}}:=\sum_{j=1}^{\infty}t^j(S_{I_{k;q}})_j\
\end{equation}
with $$(V^{(k,q-1)}_{I_{k;q}})_1=V^{(k,q-1)}_{I_{k;q}}$$ 
and, for $j\geq 2$,
\begin{eqnarray}
& &(V^{(k,q-1)}_{I_{k;q}})_j\,:=\label{formula-v_j-bis}\\
& &\sum_{p\geq 2, r_1\geq 1 \dots, r_p\geq 1\, ; \, r_1+\dots+r_p=j}\frac{1}{p!}\text{ad}\,(S_{I_{k;q}})_{r_1}\Big(\text{ad}\,(S_{I_{k;q}})_{r_2}\dots (\text{ad}\,(S_{I_{k;q}})_{r_p}(G_{I_{k;q}})\dots \Big)\\
& &+\sum_{p\geq 1, r_1\geq 1 \dots, r_p\geq 1\, ; \, r_1+\dots+r_p=j-1}\frac{1}{p!}\text{ad}\,(S_{I_{k;q}})_{r_1}\Big(\text{ad}\,(S_{I_{k;q}})_{r_2}\dots (\text{ad}\,(S_{I_{k;q}})_{r_p}(V^{(k,q-1)}_{I_{k;q}})\dots \Big)\quad\quad\quad\quad\,.
\end{eqnarray}
and, for $j\geq 1$,
\begin{equation}
(S_{I_{k;q}})_j:=ad^{-1}\,G_{I_{k;q}}\,((V^{(k,q-1)}_{I_{k;q}})^{od}_j)=\frac{1}{G_{I_{k;q}}-E_{I_{k;q}}}P^{(+)}_{I_{k;q}}\,(V^{(k,q-1)}_{I_{k;q}})_j\,P^{(-)}_{I_{k;q}}-h.c.\,.
\end{equation}
From the lines above we derive
\begin{eqnarray}
\text{ad}\,(S_{I_{k;q}})_{r_p}(G_{I_{k;q}})
&=&\text{ad}\,(S_{I_{k;q}})_{r_p}(G_{I_{k;q}}-E_{I_{k;q}})\nonumber \\
&=&\,[\frac{1}{G_{I_{k;q}}-E_{I_{k;q}}}P^{(+)}_{I_{k;q}}\,(V^{(k, q-1)}_{I_{k;q}})_{r_p}\,P^{(-)}_{I_{k;q}}\,,\,G_{I_{k;q}}-E_{I_{k;q}}]+h.c.\\
&=&-P^{(+)}_{I_{k;q}}\,(V^{(k,q-1)}_{I_{k;q}})_{r_p}\,P^{(-)}_{I_{k;q}}-P^{(-)}_{I_{k;q}}\,(V^{(k, q-1)}_{I_{k;q}})_{r_p}\,P^{(+)}_{I_{k;q}}\,.
\end{eqnarray}
We recall definition (\ref{def-S-bis}) and we observe that
\begin{equation}\label{ineq-S-V}
\|(S_{I_{k;q}})_j\|\leq 2\frac{\|(V^{(k,q-1)}_{I_{k;q}})_j\|}{\Delta_{I_{k;q}}}\leq 4\|(V^{(k,q-1)}_{I_{k;q}})_j\|\,\,,
\end{equation}
where we use the induction hypothesis that $\Delta_{I_{k;q}}\geq \frac{1}{2}$.
Then formula (\ref{formula-v_j-bis}) yields
\begin{eqnarray}
\|(V^{(k,q-1)}_{I_{k;q}})_j\|&\leq&\label{V-ineq}\\
\sum_{p=2}^{j}\,\frac{8^p}{p!}&&\sum_{ r_1\geq 1 \dots, r_p\geq 1\, ; \, r_1+\dots+r_p=j}\,\|\,(V^{(k,q-1)}_{I_{k;q}})_{r_1}\|\|\,(V^{(k,q-1)}_{I_{k;q}})_{r_2}\|\dots \|\,(V^{(k,q-1)}_{I_{k;q}})_{r_p}\|\nonumber \\
+2\|V^{(k,q-1)}_{I_{k;q}}\|&& \sum_{p=1}^{j-1}\,\frac{8^p}{p!}\,\sum_{ r_1\geq 1 \dots, r_p\geq 1\, ; \, r_1+\dots+r_p=j-1}\,\|\,(V^{(k,q-1)}_{I_{k;q}})_{r_1}\|\|\,(V^{(k,q-1)}_{I_{k;q}})_{r_2}\|\dots \|\,(V^{(k,q-1)}_{I_{k;q}})_{r_p}\|\,.\nonumber 
\end{eqnarray}
From now on,  we closely follow the proof of Theorem 3.2 in \cite{DFFR}; that is, assuming $\|V^{(k,q-1)}_{I_{k;q}}\|\neq 0$, we recursively define numbers $B_j$, $j\geq 1$, by the equations
\begin{eqnarray}
B_1&:= &\|V^{(k,q-1)}_{I_{k;q}}\|=\|(V^{(k,q-1)}_{I_{k;q}})_1\| \label{B1} \,,\\
B_j&:=&\frac{1}{a}\sum_{k=1}^{j-1}B_{j-k}B_k\,,\quad j\geq 2\,, \label{def-Bj}
\end{eqnarray}
with \,$a>0$\, satisfying the relation
\begin{equation}\label{a-eq}
\frac{e^{8a}-8a-1}{a}+e^{8a}-1=1\,.
\end{equation}
Using (\ref{B1}), (\ref{def-Bj}), (\ref{V-ineq}),  and an induction, it is not difficult to prove that (see Theorem 3.2 in \cite{DFFR}) for $j\geq 2$
\begin{equation}\label{bound-V-B}
\|(V^{(k,q-1)}_{I_{k;q}})_j\|\leq B_j\,\Big(\frac{e^{8a}-8a-1}{a}\Big)+2\|V^{(k,q-1)}_{I_{k;q}}\|\,B_{j-1}\Big(\frac{e^{8a}-1}{a}\Big)\,.
\end{equation}
From (\ref{B1}) and (\ref{def-Bj}) it also follows that
\begin{equation}\label{bound-b}
  B_j\geq \frac{2B_{j-1}\|\,V^{(k,q-1)}_{I_{k;q}}\,\|}{a}\,\quad \Rightarrow\quad B_{j-1}\leq a\frac{B_j}{2\|\,V^{(k,q-1)}_{I_{k;q}}\,\|}\,,
\end{equation} 
which, when combined with (\ref{bound-V-B}) and (\ref{a-eq}), yield
\begin{equation}\label{bound-b-bis}
B_j\geq \|\,(V^{(k,q-1)}_{I_{k;q}})_j\|\,.
\end{equation} 
The numbers $B_j$ are the Taylor's coefficients of the function
\begin{equation}
f(x):=\frac{a}{2}\cdot \left(\,1-\sqrt{1- (\frac{4}{a}\cdot \|V^{(k,q-1)}_{I_{k;q}}\|) \,x }\,\right)\,,
\end{equation}
(see  \cite{DFFR}). Therefore the radius of analyticity, $t_0$,  of 
\begin{equation}
\sum_{j=1}^{\infty}t^{j-1}\|(V^{(k,q-1)}_{I_{k;q}})^{diag}_j \|=\frac{d}{dt}\,\Big(\sum_{j=1}^{\infty}\frac{t^{j}}{j}\|(V^{(k,q-1)}_{I_{k;q}})^{diag}_j \|\Big)
\end{equation}
is bounded below by the radius of analyticity of $\sum_{j=1}^{\infty}x^jB_j$, i.e.,
\begin{equation}\label{radius}
t_0\geq \frac{a}{4\|V^{(k,q-1)}_{I_{k;q}}\|}\geq \frac{a}{8}
\end{equation}
where we have assumed $0<t<1$ and used  the assumption that $\|V^{(k,q-1)}_{I_{r;i}}\| \leq \frac{8}{(r+1)^2}\,t^{\frac{r-1}{3}}$.
Thanks to the inequality in (\ref{ineq-S-V}) the same bound holds true for the radius of convergence of the series $S_{I_{k;q}}:=\sum_{j=1}^{\infty}t^j(S_{I_{k;q}})_j\,$\,.
For $0<t<1$ and in the interval $(0,\frac{a}{16})$, by using (\ref{B1}) and  (\ref{bound-b-bis}) we can estimate
\begin{eqnarray}
\sum_{j=1}^{\infty}t^{j-1}\|(V^{(k,q-1)}_{I_{k;q}})^{diag}_j \|&\leq &\frac{1}{t}\sum_{j=1}^{\infty}t^jB_j\\
&=&\frac{1}{t}\cdot \frac{a}{2}\cdot \left(\,1-\sqrt{1- (\frac{4}{a}\cdot \|V^{(k,q-1)}_{I_{k;q}}\|) \,t }\,\right)\\
&\leq &(1+C_a \cdot t )\,\|V^{(k,q-1)}_{I_{k;q}}\|
\end{eqnarray}
for some $a$-dependent constant $C_a>0$.
Hence the inequality in (\ref{bound-V})  holds true, provided that $t$ is sufficiently small but independent of $N$, $k$, and $q$. 
In a similar way we derive (\ref{bound-S}). \qed
\end{appendix}

\end{document}